\documentclass[12pt]{article}

\usepackage[dvips]{graphics}
\usepackage{amssymb}

\newcommand{\nc}{\newcommand}
\nc{\figcap}[1]{\begin{quote}\refstepcounter{figure}
        {\bf Figure \thefigure}: {\small #1}\end{quote}}
\nc{\fig}[1]{\mbox{Fig.~\ref{#1}}}

\nc{\noi}{\noindent}

\nc{\bea}{\begin{eqnarray}}
\nc{\eea}{\end{eqnarray}}
\nc{\bean}{\begin{eqnarray*}}
\nc{\eean}{\end{eqnarray*}}
\nc{\ba}{\begin{array}}
\nc{\ea}{\end{array}}
\nc{\be}{\begin{equation}}
\nc{\ee}{\end{equation}}
\nc{\nn}{\nonumber}
\nc{\bra}[1]{\langle #1|}
\nc{\ket}[1]{|#1\rangle}
\nc{\av}[1] {\langle #1\rangle}
\nc{\vac}[1] {\langle 0| #1|0\rangle}
\nc{\amp}[2]{\langle #1|#2\rangle}
\nc{\da}{\dagger}
\nc{\pa}{\partial}
\nc{\ga}{\gamma}
\nc{\ep}{\epsilon}
\nc{\tf}{t_f}
\nc{\half}{\ensuremath{\frac{1}{2}}}
\nc{\hHH}{\hat H}
\nc{\ha}{\hat a}
\nc{\hO}{\hat O}
\nc{\hAA}{\hat A}
\nc{\hB}{\hat B}
\nc{\hG}{\hat G}
\nc{\hN}{\hat N}
\nc{\hU}{\hat U}
\nc{\hx}{\hat{x}}
\nc{\hp}{\hat{p}}
\nc{\hpsi}{\hat \psi}
\nc{\hphi}{\hat \phi}
\nc{\hpi}{\hat \pi}
\nc{\hpd}{\hat \psi ^\dagger}
\nc{\hE}{\hat E}
\nc{\hb}{\hat b}
\nc{\hc}{\hat c}
\nc{\hjo}{\hat j _0}
\nc{\hrho}{\hat \rho}
\nc{\leave}{\! \! \! \! \! / \, \,}
\nc{\intl}[1]{\int d\! #1 \,} 
\nc{\intll}[3]{\int _#1^#2 d\! #3 \,} 
\nc{\lm}{\lim _{y \rightarrow x}}

\nc{\scd}{\partial ^2 _{A_T}}
\nc{\fd}[1]{\frac{\delta }{\delta #1}} 
\nc{\pad}[1]{\frac{\partial}{\partial #1}} 
\nc{\refpa}[1]{(\ref{#1})} 

\nc{\calH}{\ensuremath{\mathcal{H}}}
\nc{\calD}{\ensuremath{\mathcal{D}}}
\nc{\calL}{\ensuremath{\mathcal{L}}}
\nc{\calO}{\ensuremath{\mathcal{O}}}
\nc{\hcalO}{\ensuremath{\hat \mathcal{O}}}
\nc{\calK}{\ensuremath{\mathcal{K}}}
\nc{\Tr}{\ensuremath{\mathrm{Tr}}}
\nc{\tr}{\ensuremath{\mathrm{tr}}}
\nc{\ra}{\rightarrow}
\nc{\lr}{\leftrightarrow}
\nc{\phistar}{\phi^*}
\nc{\etat}{\eta_T}
\nc{\het}{\hat E_T}
\nc{\hpt}{\hat \psi_T}
\nc{\hpdt}{\hat \psi ^\dagger_T}
\nc{\bart}{\bar{t}}
\nc{\barp}{\bar{p}}
\nc{\barT}{\bar{T}}
\nc{\hbarrho}{\hat{\bar{\rho}}}
\nc{\bga}{\ensuremath{\mbox{\boldmath{$\gamma$}}}}
\nc{\bsi}{\ensuremath{\mathbf{\sigma}}}
\nc{\bx}{\ensuremath{\mathbf{x}}}
\nc{\by}{\ensuremath{\mathbf{y}}}
\nc{\bz}{\ensuremath{\mathbf{z}}}
\nc{\bp}{\ensuremath{\mathbf{p}}}
\nc{\bn}{\ensuremath{\mathbf{n}}}
\nc{\bbp}{\ensuremath{\bar{\mathbf{p}}}}
\nc{\bP}{\ensuremath{\mathbf{P}}}
\nc{\hbA}{\hat{\ensuremath{\mathbf{A}}}}
\nc{\hbB}{\hat{\ensuremath{\mathbf{B}}}}
\nc{\bA}{\ensuremath{\mathbf{A}}}
\nc{\bJ}{\ensuremath{\mathbf{J}}}
\nc{\bB}{\ensuremath{\mathbf{B}}}
\nc{\bH}{\ensuremath{\mathbf{H}}}
\nc{\bM}{\ensuremath{\mathbf{M}}}
\nc{\bD}{\ensuremath{\mathbf{D}}}
\nc{\bE}{\ensuremath{\mathbf{E}}}
\nc{\hbE}{\hat{\ensuremath{\mathbf{E}}}}
\nc{\br}{\ensuremath{\mathbf{r}}}
\nc{\bj}{\ensuremath{\mathbf{j}}}
\nc{\bOm}{\ensuremath{\mathbf{\Om}}}
\nc{\om}{\omega}
\nc{\Om}{\Omega}
\nc{\sgn}{\mbox{sgn}}
\nc{\deltabar}{\mbox{$\delta\hspace*{-8pt}\vspace*{-8pt}-$}}
\nc{\gammat}{\tilde{\gamma}}
\nc{\binom}[2] {{#1\choose #2}}
\nc{\mub}{\bar{\mu}}
\nc{\rhob}{\bar{\rho}}
\nc{\Bb}{\bar{B}}
\nc{\Jb}{\bar{J}}
\nc{\Mb}{\bar{M}}
\nc{\Tb}{\bar{T}}
\nc{\sbar}{\bar{s}}
\nc{\betab}{\bar{\beta}}
\nc{\hj}{\hat j}
\nc{\hQ}{\hat Q}
\nc{\hJ}{\hat J}
\nc{\hA}{\hat A}
\nc{\hH}{\hat H}
\nc{\de}{\delta}
\nc{\leri}{\leftrightarrow}
\nc{\llabel}[1]{\label{#1}\marginpar{#1}}

\nc{\bc}{\begin{center}}
\nc{\ec}{\end{center}}
\nc{\inv}[1]{\frac{1}{#1}}

\newlength{\overeqskip}
\newlength{\undereqskip}
\nc{\eq}[1]{\mbox{Eq.~(\ref{#1})}}
\nc{\eps}{\epsilon}
\nc{\goto}{\rightarrow}

\nc{\cF}{{\cal F}}
\nc{\cG}{{\cal G}}
\nc{\cH}{{\cal H}}

\begin{document}
%
%
%
%
%
%
\thispagestyle{empty} 
%
%
%
%
{\begin{tabular}{lr} 
%
%
\, 
& 
~~~~~~~~~~~~~~~~~~
~~~~~~~~~~~~~~~~~~~~~~~~~~~~~~~~~~~~~~~~~~~~~~~~~~~~~~~~~CERN-TH/99-410 %
\end{tabular}} 
%
%
\vspace{10mm} 
\begin{center} 
\baselineskip 1.2cm 
{\Huge\bf   Theory of the Microscopic Maser Phase Transitions 
}\\[1mm] 
\normalsize 
\end{center} 
{\centering 
{\large Per K. Rekdal\footnote{Email address: perr@phys.ntnu.no.}$^{,a}$ 
and 
{\large Bo-Sture K. Skagerstam\footnote{Email 
address: boskag@phys.ntnu.no.}$^{,a,b}$} 
 \\[5mm] 

$^{a}~$Department of Physics, 
The Norwegian University of Science and Technology,   
N-7034 Trondheim, Norway \\[1mm]
$^{b}~$Theory Division, CERN, CH-1211 Geneva 23, Switzerland\\[1mm]
} } 
%
%
%
%
\begin{abstract} 
\normalsize 
%
%
\vspace{5mm}

\noindent    
    Phase diagrams of the micromaser system are mapped out in terms of the physical 
    parameters at hand like the atom cavity 
    transit time $\tau$, the atom-photon frequency detuning $\Delta \omega$,
    the number of thermal photons $n_b$ and the probability $a$ for a pump
    atom to be in its excited state. 
    Critical fluctuations are studied in terms of correlation measurements on 	atoms having 
    passed 
    through the micromaser or on the microcavity photons themselves.
    At sufficiently large values of the detuning we find a ``twinkling''
    mode of the micromaser system. Detailed properties of the trapping states 	are
    also presented.
    
\end{abstract}

\newpage

\bc{
\section{INTRODUCTION}
}\ec

  	There are few systems in physics which exhibit a rich structure  of phase 	transitions that can be investigated under clean experimental conditions 	and  which  at 	the 	same time  can be studied by exact theoretical 	methods. 
  	Recent developments in cavity quantum electrodynamics, and in particular 	the 	use of beams of highly excited Rydberg atoms and resonant 	microcavities, has opened up an avenue to 
  	study one class of such systems in great detail. 
  	The micromaser system, which is a remarkable experimental realization of 	the 	idealised system of a two-level  atom interacting with a 
	second-quantised
  	single-mode electromagnetic  field, provides  us with such an example  
  	(for reviews and references see e.g. \cite{Walther88}).  
  	The microlaser \cite{An94} is the counterpart in the optical regime.

  Many features of the micromaser are of general interest. It can, for example, be argued
  that  the micromaser system is a very simple illustration of the conjectured  
  topological origin of second-order phase transitions \cite{Casetti99},
  as will be briefly discussed later in the present paper.
  Various aspects of stochastic resonance can, furthermore, be studied in this   system
  \cite{Buchleitner98}. The micromaser also
  illustrates a remarkable feature  of non-linear dynamical systems: turning on
  randomness may lead to an increased  signal-to-noise ratio \cite{noise94,seoul&99}.

   The basic theory of the micromaser system
   as  developed in Refs.\cite{Filipowicz86,Guzman89},
   which is limited to the case $a=1$ and $\Delta \omega =0$ only, 
   suggests the existence of various phase transitions 
   in the large $N$ limit. 
   Here $N$ is the number of atoms passing the cavity in a single photon
   decay time.   
   A natural order parameter
   is then the average photon ``density'' $\langle n\rangle /N$, 
   where $\langle n\rangle$ denotes the average photon number with
   respect to the stationary micromaser photon distribution. 
   By making use of an
   approximative approach in terms of a  Fokker-Planck equation, 
   details of the various transitions were worked out, like tunnelling times
   between the various phases (see e.g. \cite{Filipowicz86,Guzman89}). 
   An exact treatment, in the large $N$ limit,
   of the micromaser
   phase structure and the corresponding critical fluctuations in
   terms of a conventional correlation function of either the atoms
   leaving the micromaser or the photons in the cavity has been given in
   Refs.\cite{ElmforsLS95}, where deviations from the Fokker-Planck equation approach were reported. It is of great interest to notice that
   spontaneous jumps in  $\langle n\rangle /N$ 
   and  large correlation lengths close to micromaser phase transitions have
   been observed experimentally \cite{Walther88,Walther97}.
   Most of the experimental and theoretical 
   studies have, however, been limited to the case $a=1$ and
   $\Delta \omega = 0$. It is the purpose of the present paper to study the
   phase structure of the micromaser system and extend the results of
   Refs.\cite{ElmforsLS95} to the  general case with $a\neq 1$ and
   $\Delta \omega \neq 0$ using methods which are exact in the large $N$ limit. 
   As will be argued below, several new
   intriguing physical properties of the micromaser system are then unfolded. A short
   presentation of the main results of this paper has appeared 
   elsewhere \cite{Rekdal&Skagerstam&99a}.

   The paper is organised as follows. In Section \ref{p_distr_KAP} the theoretical
   framework is outlined, and a rapidly convergent
   large $N$ limit is derived for the stationary
   micromaser photon probability distribution. The phase structure is investigated in 
   Section \ref{phase_diagr_KAP} and the order parameter $\langle x \rangle$
   is discussed in more detail in Section \ref{order_param_KAP}.
   The correlation length is analyzed numerically in Section \ref{corr_KAP} as well
   as in terms of various approximation schemes. The results
   of extensive numerical investigations of trapping states are presented in Section
   \ref{trapping_Sec}. Final remarks are given in \ref{final_KAP}.

\vspace{1.0cm} 
\bc{
\section{THE PROBABILITY DISTRIBUTION}
\label{p_distr_KAP}
}\ec

   The pump atoms, which enter the cavity, are assumed to be prepared in an 
   incoherent statistical mixture, i.e. the initial density matrix $\rho_A$ of the atoms 
   is of the diagonal form

\be \label{pho_A} 
   \rho_A = 
   \left  (
          \ba{rr}
           a & 0  \\
           0 & b
          \ea 
   \right )~~,
\ee

   \noi
   where $a+b=1$. This form of the atomic density matrix is not of the most general form, but it 
   leads to the possibility of an exact analytical treatment provided the photon field density 
   matrix also is diagonal, which from now on we assume. The rate $R$ of the injected atoms is 
   assumed to be high enough to pump up the cavity, 
   i.e. $R>\gamma =1/T_{cav}$, or $N>1$ in terms of the 
   dimensionless flux variable $N=R/\gamma$. Here $T_{cav}$ is the
   decay time of the cavity and hence $\gamma$ is the typical decay rate for 
   photons in the cavity. Furthermore, let $\tau$ denote the atomic transit-time through the 
   resonant photon cavity.

   For our purposes the continuous time formulation of the micromaser system 
   \cite{Lugiato87} is suitable as a theoretical framework.
   Each atom is assumed to have  the same probability $R \, dt$ of 
   arriving in an infinitesimal time interval $dt$,
   i.e. they are Poisson distributed.
   Provided the interaction 
   with the radiation field of the cavity takes less time than this interval, i.e. if $\tau \ll dt$, we 
   may consider the atomic transitions to be instantaneous.
   The vector $p$ formed by the diagonal density matrix elements of 
   the photon field then obeys the master equation \cite{Lugiato87,ElmforsLS95}

\be \label{master_eq} 
   \frac{dp}{dt} = - \gamma L p~~,   
\ee 

   \noi
    where 

\be \label{L_matise}
  L = L_C - N(M-1) ~~.
\ee

   \noi
   Here $L_C$ describes the damping of the cavity
   (see e.g.  Ref. \cite{Scully&Zubairy} for an excellent account), i.e.

\be
  (L_C)_{nm} = (n_b+1)[ \, n \delta_{n,m} - (n+1) \delta_{n+1,m} \, ] + 
              n_b[ \, (n+1)\delta_{n,m}
              - n \delta_{n,m+1} \, ]~~,
\ee

   \noi
   and $M = M(+) + M(-)$ has its origin in the Jaynes-Cummings (JC) 
   model \cite{Jaynes63,ElmforsLS95} where

\be \label{M_plus}
   M(+)_{nm} = b q_{n+1} \delta_{n+1,m} + a(1-q_{n+1}) \delta_{n,m}~~,
\ee

\noindent and
\be \label{M_minus}
   M(-)_{nm} = a q_{n} \delta_{n,m+1} + b(1-q_{n}) \delta_{n,m}~~.
\ee

  \noi
  We observe the unitarity constraints

\bea \label{sums_eq}
  \sum_{n=0}^{\infty}M_{nm} = 1~~,\nonumber\\
  \sum_{n=0}^{\infty}(L_C)_{nm} = 0~~.
\eea

   \noi
   The stationary solution of \eq{master_eq} is well known \cite{Filipowicz86,Lugiato87}
   and is given by

\be \label{p_n_eksakt}
   {\bar p}_n = {\bar p}_0 \prod_{m=1}^{n} \frac{n_b \, m +  N a q_m}
         {(1+n_b) \, m + N b q_m}~~, 
\ee
   where $q_m \equiv q(m/N)$ and 
\be \label{q}
    q(x) = \frac{x}{x + \Delta^2}~
           \sin^2\left(\theta
           \sqrt{ x + \Delta^2 }~ \right)~~.
\ee

   \noi
   Here we have defined the natural and dimensionless parameter

\be
   \Delta^2 = \delta^2/N~~,
\ee

   \noi where

\be
   \delta = \Delta \omega /2g~~,
\ee

   \noi
   since it is always this particular combination of $\Delta \omega$, $N$ and $g$ 
   which is involved. $g$ is the single-photon Rabi frequency at zero
   detuning of the JC model \cite{Jaynes63}. $n_b$ is the initial mean value of thermal photons in the cavity.
   Furthermore, the probability is expressed in terms of the scaled dimensionless pump 
   parameter $\theta = g\tau \sqrt{N}$.
   The overall constant ${\bar p}_0$ in \eq{p_n_eksakt} is determined by the
   normalisation condition

\be
   \sum _{n=0}^{\infty} {\bar p}_n =1~~.
\ee

   \noi
   In passing, we observe that at thermal equilibrium, i.e. when

\be \label{a_b_equilibr}
   \frac{a}{b} = \frac{n_b}{1+n_b}~~,
\ee

   \noi
   the probability distribution in \eq{p_n_eksakt} is in fact, as it should be,
   a Planck distribution. This particular distribution depends only 
   on $n_b$, which is the corresponding mean value photon number,
   i.e. 

\be
   \langle n \rangle \equiv \sum_{n=0}^{\infty} \,n {\bar p}_n  = n_b~~.
\ee

   \noi
   If \eq{a_b_equilibr} is not fulfilled then ${\bar p}_n$ describes a stationary
   photon distribution which therefore can be far from thermal equilibrium.

\vspace{0.5cm}
\bc{
\subsection{Large N Expansion}
}\ec

  The large $N$ behaviour of the stationary photon probability distribution ${\bar p}_n$ as given by \eq{p_n_eksakt} is difficult to handle.
   In order to obtain 
   an appropriate  expression   
   which is rapidly convergent in the 
   large $N$ limit,
   and which neatly expresses the various micromaser phases, 
   we can make use  of a Poisson summation technique \cite{Poisson}. 
   It is then natural to define the scaled photon number variable $x$, 
   defined by $x = n/N$. The stationary probability distribution \eq{p_n_eksakt} 
   then takes the form

\be \label{p_x}
   {\bar p}(x) = {\bar p}_0 \sqrt{\frac{w(x)}{w(0)}} ~ e^{-N \, V(x)}~~,
\ee

   \noi where 

\be
   V(x) = \sum_{k=-\infty}^{\infty}V_k(x)~~. 
\ee

\noi The effective potential $V(x)$ is expressed in terms of



\be \label{V_k}
   V_k(x) = -  \int _0^x d\nu  \, \ln[\, w(\nu)\,] \, \cos(2\pi N k \nu) ~~, 
\ee

  \noi where

\be \label{w_x}
   w(x) = \frac{n_b \, x + a \, q(x)}{(1+n_b)\, x + b \, q(x)}~~.
\ee

   \noi
   In this expression for $w(x)$, $q(x)$ is given by \eq{q}.  
   We stress that \eq{p_x} is exact.

   In the large $N$ limit \eq{p_x} can be substantially simplified since
   the $V_k(x)$-terms in \eq{V_k} do not contribute in this limit for 
$k \neq 0 $ .
   It is therefore the nature of the global minima of $V_0(x)$ which determines 
   the probability distribution and the micromaser phase structure, apart from the 
   zeros of $w(x)$ which correspond to trapping 
   states \cite{Fili&Java&Meystre}-\cite{W&V&H&W}. The physics of trapping 
   states will be discussed in Section~\ref{trapping_Sec}.

\vspace{0.5cm}
\bc{
\subsection{The Thermal Distribution}
}\ec

   If the only global minimum of $V_0(x)$ occurs at $x=0$, we can expand 
   the effective potential $V_0(x)$ around the origin. 
   A straightforward expansion of $V_0(x)$ then leads to

\be \label{V_0_utv}
  V_0(x) = x \ln \left [ \frac{1+n_b+b \theta^2_{eff}}{ n_b+ a\theta^2_{eff} } \right ] + 
{\cal O}(x^2) ~~,
\ee

  \noi
  where we have defined an effective pump $\theta$-parameter

\be
   \theta_{eff}^2 = \theta^2 \, \frac{\sin^2(\theta \Delta)}{(\theta \Delta)^2}~~.
\ee

  \noi
   \eq{V_0_utv} leads to the following properly normalised thermal 
photon probablity distribution

\be \label{p_n_geo}
   {\bar p}_n = \frac{1+(1-2a)\,\theta_{eff }^2}{1+n_b + b\,\theta_{eff}^2} 
               \left ( \frac{n_b+a\,\theta_{eff}^2}{1+n_b+b\,
               \theta_{eff}^2} \right )^n~~,
\ee

   \noi
   which is convergent provided

\be \label{conv_cond}
   \theta_{eff}^2 \, (2a-1)<1~~.
\ee

   \noi
   \eq{p_n_geo} is  exact in the large $N$ limit, and corresponds to an 
   increase of the temperature in the cavity.
   If $a<1/2+\Delta^2/2$ the distribution \eq{p_n_geo} is therefore always valid. 
   If $\Delta=0$ convergence requires that  $\theta^2 \, (2a-1)<1$, 
   i.e. $\theta$ has to be sufficiently small when $a>1/2$.
   We find that \eq{p_n_geo} also can be used to compute expectation values for 
   finite $N$, even though the accuracy then depends on the actual micromaser 
   parameters.
   When the micromaser is described by the thermal distribution in \eq{p_n_geo},
   we say that the system is in the thermal phase.

\vspace{0.5cm}
\bc{
\subsection{The Gaussian Distribution}
}\ec

   If, on the other hand, non-trivial saddle-points of $V_0(x)$ exist,
   \eq{p_x} can be written in the form

\be
  {\bar p}(x) = \sum_j {\bar p}_j(x) ~~,
\ee

   \noi
   where $\sum_j$ is a sum over all saddle-points corresponding to minima of $V_0(x)$,
   and where ${\bar p}_j(x)$ is ${\bar p}(x)$ for $x$ close to the saddle-points $x=x_j$.
   The saddle-points, $x(\theta)$, are determined by $V_0^{\prime}(x)=0$, 
   since, as noted above, the $V_k(x)$-terms in \eq{V_k} for $k \neq 0$ do not contribute in the 
   large  $N$-limit. For a fixed $a$ and $\Delta$ we then have

\be \label{trans}
   (2a-1) \, \sin^2\left(\theta \sqrt{ x(\theta) + \Delta^2 }~ \right)  -  
             (x(\theta) + \Delta^2)  = 0 ~~,
\ee

   \noi
   corresponding to $w(x(\theta))=1$.
   This equation is independent of $n_b$ and has non-trivial solutions 
   only when $a\geq1/2+\Delta^2/2$.
   It is clear that $a$, $\Delta$ and $x(\theta)$ at a saddle-point are restricted by the 
   condition

\be
   x(\theta)+\Delta^2 \leq 2a-1 ~~,
\ee

   \noi
   for any $\theta$.
   The saddle-points $x=x(\theta)$
   can conveniently be parametrically represented in the 
   form \cite{ElmforsLS95}

\be \label{x_phi}
   x(\phi) + \Delta^2 = (2a-1) \sin^2 \phi ~~,  
\ee

\be \label{theta_phi}
   \theta(\phi) = \frac{1}{\sqrt{2a-1}} \frac{\phi}{|\sin \phi |}~~,
\ee

   \noi
   with 

\be  \label{phi_0}
  \phi \geq \phi_0 \equiv \arcsin(|\Delta|/\sqrt{2a-1})~~.
\ee

   \noi
   We define branches of $\phi$, labelled by $k=0,1,2,...$ such that $\phi$
   varies in the intervals

\be \label{PHI_interv} 
   \phi_0 +k\pi \leq \phi \leq (k+1)\pi -\phi_0~~,~~k=0,1,2,...~~. 
\ee

  \noi
  Except for the first branch $k=0$, the saddle-points corresponding to each of these 
  branches are double-valued, that is, there are at most two values of $x(\theta)$
  for a given $\theta$. The first branch $k=0$ is single-valued.

   For a given $a$, $\Delta$ and $\theta$, let $x=x_j(a,\Delta, \theta) \equiv x_j$ 
   be a saddle-point corresponding to a minimum of $V_0(x)$.  
   The equilibrium distribution in the neighbourhood of this minimum 
   value $x_j$ is then given in the following Gaussian form

\be \label{p_j}
   {\bar p}_j(x) = \frac{T_j}{\sqrt{2\pi N}}
                   ~e^{-\frac{N}{2} V_0^{\prime \prime}(x_j)(x-x_j)^2}~~.
\ee

  \noi 
  Here $T_j$ is determined by the normalisation condition for ${\bar p}(x)$, 
  i.e

\be  \label{T_j}
   T_j = \frac{e^{- N V_0(x_j)}}{ {\displaystyle \sum_m} 
         \displaystyle{ e^{-N V_0(x_m)}}/\sqrt{V_0^{\prime \prime}(x_m)}}~~,
\ee

  \noi and where

\be \label{V_dobbel}
   V_0^{\prime \prime}(x) =  
\frac{(2a-1)^2}{a + n_b(2a-1)} \, \frac{q(x) - x q^{\prime}(x)}{x^2} ~~.
\ee

   \noi
   For the given parameters, the sum in \eq{T_j} is supposed to be taken 
   over all saddle-points corresponding to a minimum of $V_0(x)$, 
   i.e. all saddle-points corresponding to $V_0^{\prime \prime}(x)>0$.
   If $x=x_j$ does not correspond to a global minimum 
   for the effective potential $V_0(x)$,  
   then $T_j$ is exponentially small in the large $N$ limit.
   If $x=x_j$ corresponds to one and only one global 
   minimum,  we can neglect all 
   the terms in the sum in $T_j$ but $m=j$, in which case $T_j$ is 
   reduced to  $T_j = \sqrt{V_0^{\prime \prime}(x_j)}$. 
   In the neighbourhood of such a global minimum $x_j$ the probability distribution in 
   \eq{p_j} is therefore reduced to

\be \label{p_j_gauss}
   {\bar p}_j(x) = \sqrt{\frac{V_0^{\prime \prime}(x_j)}{2\pi N} }~ 
                   e^{-\frac{N}{2}{V_0^{\prime \prime}(x_j)}(x-x_j)^2}~~.
\ee

   \noi 
   which is, strictly speaking, only valid in a sufficiently small neighbourhood 
   of $x_j$.

   However, since ${\bar p}_j(x)$ in \eq{p_j} is exponentially small when $x$ is not in the 
   neighbourhood of $x_j$,  the probability distribution for any $x$ is 
   given by

\be \label{p_x_approx}  
  {\bar p}(x) =  \sum_{j^*} {\bar p}_j(x)~~,
\ee

   \noi
   where $\sum_{j^*}$ denotes the sum over the global minima of $V_0(x)$
   only. Hence, when there are several saddle-points, the actual maser phase is 
   described by the saddle-points which correspond to the global minima
   of $V_0(x)$ only. 
   In the next section, we will actually argue that, for a fixed $a$, $\Delta$ 
   and $\theta$, we can have at most two competing global minima
   in the large $N$ limit,
   i.e. there can be at most two terms in the $\sum_{j^*}$-sum in \eq{p_x_approx}.

   When the micromaser is described by a distribution like \eq{p_j_gauss} the mean occupation 
   number $\langle n \rangle$ is proportional to the
   pumping rate $N$ (see Section~\ref{order_param_KAP}). 
   The cavity then acts as a maser, i.e. the system is in a maser phase.

\vspace{1cm} 
\bc{ 
\section{THE PHASE DIAGRAM}
\label{phase_diagr_KAP}
}\ec

   The probability distribution \eq{p_n_eksakt} determines 
   micromaser phases as a function of the physical parameters at hand.
   In this section we will map out  
   phase diagrams in the $a$- and $\theta$-parameter space for a given $n_b\neq0$
   and $\Delta$.
   In general a phase diagram is then determined 
   by mapping out the global minima 
   of the effective potential $V_0(x)$.

   By the substitution $\phi = \theta \sqrt{x+\Delta^2}$, the effective potential 
   in \eq{V_k} can be written in the form

\be \label{V_phi}
   V_0(x) = V_0(\phi,\theta) = - \frac{2}{\theta^2} \, 
                \int _{\theta |\Delta|}^{\phi} d\chi \, \chi \, 
                \ln[\, w(\chi,\theta)\,] ~~.
\ee

  \noi Here we have defined 

\be \label{w_phi}
   w(\chi,\theta) = \frac{n_b + a \, q(\chi,\theta)}{1+n_b + b\, q(\chi,\theta)}~~,
\ee

  \noi and

\be \label{q_phi}
    q(\chi,\theta) =  \theta^2 ~ \frac{ \sin^2 \chi }{ \chi^2  }~~.
\ee

   \noi
   In \eq{V_phi}, the upper integration limit $\phi$ and the pump parameter $\theta$
   may take on arbitrary values.
   However, by choosing $x(\phi)$ and $\theta(\phi)$ according to \eq{x_phi} and \eq{theta_phi}
   the effective potential is always at an extremum. 
   We then arrive at an effective multi-branched potential as given by
   $V_0(\phi) = V_0(\phi, \theta(\phi))$.
   For a given branch $k$, the variable $\phi$ is limited by \eq{PHI_interv}.

   If we regard $V_0(\phi)$ as a function of $\theta$, each branch of 
   $V_0(\theta)=V_0(\phi,\theta(\phi))$ 
   is at most double-valued  except for branch $k=0$.
   One sub-branch then corresponds to a maximum 
   ($V_0^{\prime \prime}(x(\phi))<0$), which is swept out first as $\phi$ increases,
   and the other corresponds to a minimum ($V_0^{\prime \prime}(x(\phi))>0$).
   Here the $V_0^{\prime \prime}(x(\phi))$ is explicitly given by

\be \label{V_dobbel_PHI}
   V_0^{\prime \prime}(x(\phi)) =  \frac{1 - \phi  \cot  \phi }
                              { \sin ^2 \phi \,[a+n_b(2a-1)]} ~~.
\ee

\noi   
   To reconstruct the actual value of $V_0(x)$ for the $k$:th minimum (maximum), we write
   $V_0(x)=V_0(\phi)$, evaluated for the sub-branch with $V_0^{\prime \prime}(x(\phi)) > 0$
   ($V_0^{\prime \prime}(x(\phi)) < 0$).

   A convenient representation of $V_0(\phi)$ for the branch $k$ (given $a$, $n_b$, $\Delta$) is

\be \label{V_min}
   V_0(\phi) = [a+n_b(2a-1)] \, I(\phi) ~~,
\ee

  \noi where

\be
  I(\phi) \equiv \int_{\phi_0+k\pi}^{\phi} ~ d\chi \,
                 \chi^4 \,
                  \frac{\sin(2 \chi) - 2 \sin^2 (\chi)/\chi }
                       { \left [ \chi^2 n_b + a \, \theta(\phi)^2 \sin^2 \chi \right ] 
                         \left [ \chi^2(1+n_b) + b \, \theta(\phi)^2 \sin^2\chi \right ] } ~~. 
\ee

  \noi 
  Here the upper integration limit $\phi$ has to be chosen according to \eq{PHI_interv}. 
  We notice the characteristic pre-factor $a+n_b(2a-1)$ in \eq{V_min} which actually determines the typical dependence of the parameters $a$ and $n_b$ at a saddle-point.

\vspace{0.5cm}
\bc{
\subsection{The Critical Parameters $\theta_0^*$ and $\theta_k$}
}\ec

   As a function of $\theta$, 
   the first extremum of $V_0(x)$ occurs at $\theta_0^* \equiv \theta(\phi_0)$,
   where  $\phi_0$ is given by \eq{phi_0}. The critical pump 
   parameter $\theta_0^*= \theta_0^*(a,\Delta)$
   is therefore given by

\be \label{theta_0_stj}
  \theta_0^* = \frac{\arcsin(|\Delta|/\sqrt{2a-1})}{|\Delta|} ~~, 
\ee

   \noi
   which is equivalent to

\be \label{a_sfa_theta}
   a = \frac{1}{2} + \frac{\Delta^2}{2 \sin^2(\theta_0^* \Delta)} ~~.
\ee

   \noi  
   When the system is not detuned, i.e. when $\Delta=0$,
   this equation reduces to

\be \label{a_sfa_theta_0}
   a = \frac{1}{2} + \frac{1}{2 (\theta_0^*)^2}~~.
\ee

 \noi
   \eq{theta_0_stj} determines the first thermal-maser critical line in the 
   $a$- and $\theta$- phase diagram.
   In the maser phase, 
      the order parameter $\langle x \rangle = \langle n \rangle/N$ approaches zero when
   $\theta$ approaches $\theta_0^*(a,\Delta)$.
   Furthermore,
   in the large $N$ limit, $\langle x \rangle$ is always zero in 
   the thermal phase (see Section~\ref{order_param_KAP}). 
   Hence, the order parameter is continuous on this critical line.
   The first derivative of $\langle x \rangle$ with respect to
   $\theta$ is, however, discontinuous. In the maser phase
   we actually have

\be \label{der_x_theta}
  \frac{d \langle x \rangle}{d \theta} = 2 (\sqrt{2a-1})^3 \, 
\frac{\sin^3 \phi}{\tan \phi - \phi} ~~,
\ee

   \noi
   with $\phi \geq \phi_0$.
   When approaching the critical line from the maser phase, we see from \eq{der_x_theta} that 
   $d \langle x \rangle /d \theta$ is non-zero (except for $a=1/2$).
   The first thermal-maser critical line therefore corresponds to 
    a line of  second-order phase transitions.

   In passing, we mention that in a  topological analysis of the second-order phase transitions 
   \cite{Casetti99} of the micromaser system, $V_0(x)$ will play the role of a Morse function.
   The configuration space ${\cal M}$ is then the one-dimensional space of $x(\equiv n/N)$. The 
   sub-manifold ${\cal M}_v$ defined by

\be
  {\cal M}_v = \{ x \in {\cal M} | ~ {\bar p}(x) \simeq \exp( \, -N V_0(x) \, ) \leq v \} ~~,
 \ee

  \noi
  then describes the change of topology  close to the second-order phase transitions 
  $\theta_0^*$. This will be true for all second-order  transitions discussed below.
 We observe that for finite N the change 
  in topology, i.e. the appearance of an additional disjoint
  $x$-interval due to the appearance of a new
  local maximum in ${\bar p}(x)$, in general 
  occurs before the actual second-order phase transition.

\begin{figure}
\unitlength=1mm
\begin{picture}(100,80)(0,0)

\includegraphics{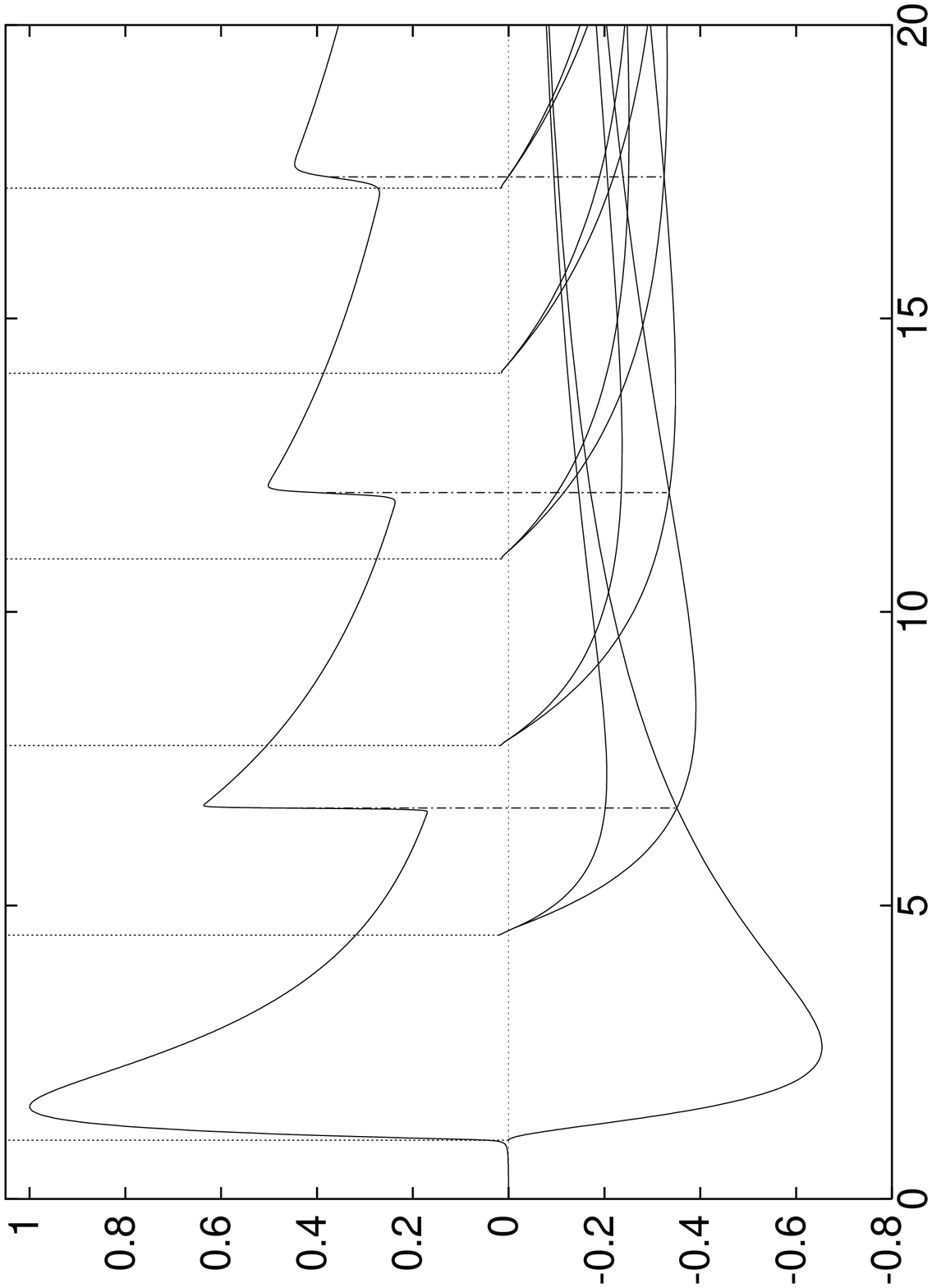}      

   \put(87,-6){\normalsize  \boldmath$\theta$}
   \put(-2,25){\normalsize  \boldmath$V_0(\theta)$}
   \put(-2,74){\normalsize  \boldmath$\langle x \rangle$}

   \put(26.8,106){\normalsize  \boldmath$\theta_0^*$}
   \put(50.8,106){\normalsize  \boldmath$\theta_1$}
   \put(72.2,106){\normalsize  \boldmath$\theta_2$}   
   \put(93.2,106){\normalsize  \boldmath$\theta_3$}   
   \put(114,106){\normalsize  \boldmath$\theta_4$}   
   \put(135,106){\normalsize  \boldmath$\theta_5$}

   \put(65.7,27.12){\circle{2.7}}
   \put(64.7,20.12){\normalsize  \boldmath$\theta^*_{01}$}

   \put(101.55,27.8){\circle{2.7}}
   \put(100.55,20.8){\normalsize  \boldmath$\theta^*_{12}$}
 
   \put(137.4,28.33){\circle{2.7}}
   \put(136.4,21.33){\normalsize  \boldmath$\theta^*_{23}$}

\end{picture}
\vspace{8mm}
\figcap{ The order parameter $\langle x \rangle = \langle n \rangle /N$, 
         as a function of $\theta$, when $a=1$, $\Delta=0$, $n_b=0.15$ and $N=1000$.
         A dotted vertical  line at
         $\theta=\theta_k$, where $\theta_0 = \theta_0^*$, 
         indicates where the $k$:th branch comes 
         into existence. $\theta_0^*$ corresponds to a second-order phase transition.
         The numerical values of the pump parameters are: 
         $\theta_0^*=1$, $\theta_1 = 4.603$, $\theta_2=7.790$, $\theta_3 = 10.950$,
         $\theta_4 = 14.102$ and $\theta_5 = 17.250$.
         The various branches at the extremum of the
         effective potential, i.e. $V_0(\theta)=V_0(\phi ,\theta(\phi))$, 
         are also shown.
         The intersection between two neighbouring $V_0^{\prime \prime}(x(\phi))>0$ 
         sub-branches at $\theta=\theta^*_{k k+1}$
         is marked by a circle when this coincidence occurs at a 
         global minimum of $V_0(\theta)$.
         At these first-order maser-to-maser phase transitions  
         the order parameter $\langle x \rangle$ 
         makes a discontinuous change in the large $N$ limit.
         The numerical values of the transition parameters $\theta^*_{k k+1}$
         are: $\theta^*_{01}\approx6.6610$, 
         $\theta^*_{12}\approx12.035$ and  $\theta^*_{23}\approx17.413$.
\label{x_og_V_Delta0} }
\end{figure}

\begin{figure}
\unitlength=1mm
\begin{picture}(100,80)(0,0)

\includegraphics{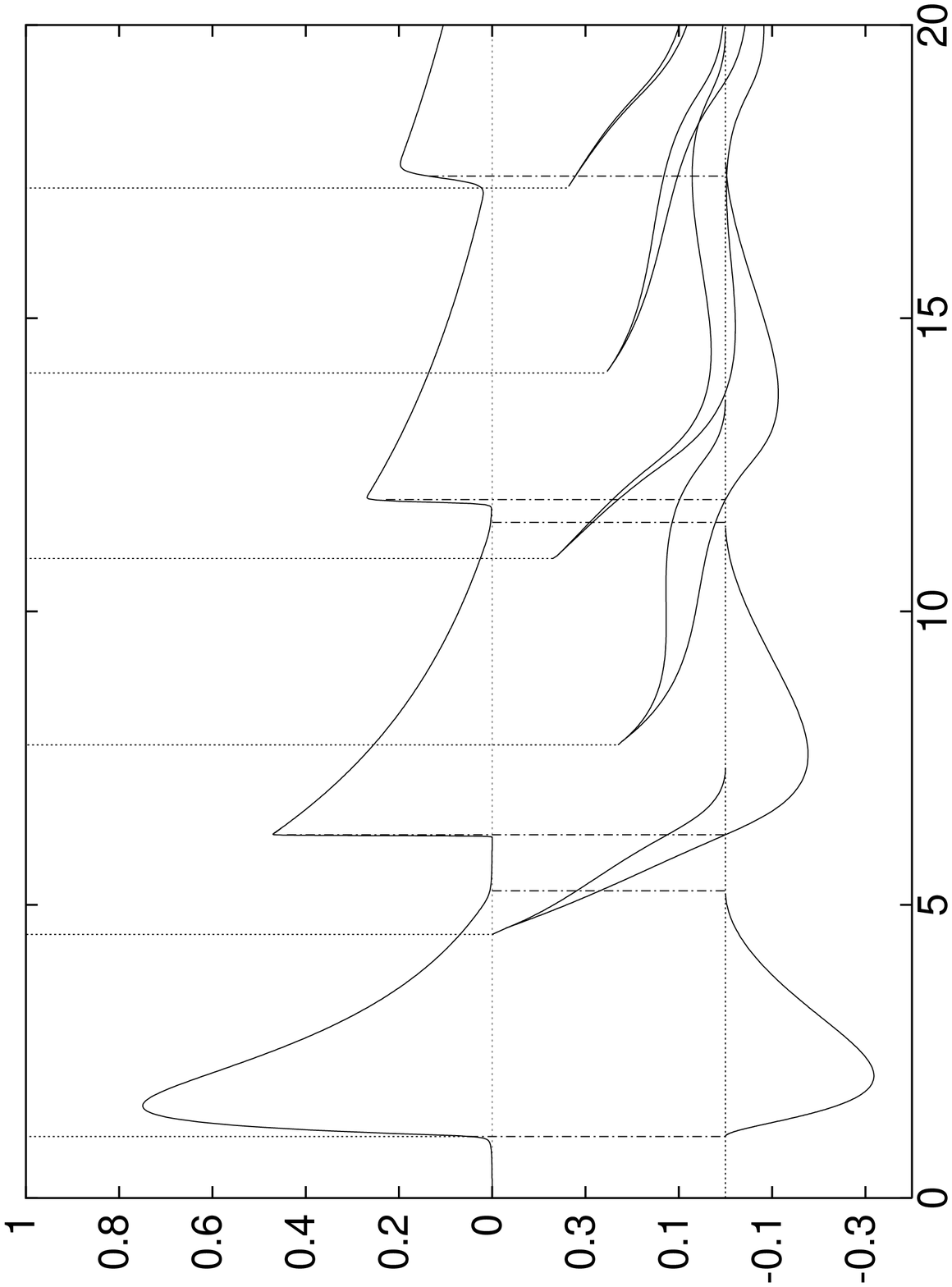}      

   \put(87,-8){\normalsize  \boldmath$\theta$}
   \put(-2,25){\normalsize  \boldmath$V_0(\theta)$}
   \put(-2,75){\normalsize  \boldmath$\langle x \rangle$}
   
   \put(137.5,23.7){\circle{2.7}}   

   \put(26.8,106){\normalsize  \boldmath$\theta_0^*$}
   \put(50.8,106){\normalsize  \boldmath$\theta_1$}
   \put(72.2,106){\normalsize  \boldmath$\theta_2$}   
   \put(93.2,106){\normalsize  \boldmath$\theta_3$}   
   \put(114,106){\normalsize  \boldmath$\theta_4$}   
   \put(135,106){\normalsize  \boldmath$\theta_5$}

   \put(100.97,14){\vector(0,1){10}}
   \put(103.45,14){\oval(5,5)[bl]}

   \put(98.3,14){\vector(0,1){10}}
   \put(95.82,14){\oval(5,5)[br]}

   \put(88,10.5){\normalsize  \boldmath$\theta_{1t}^*$}
   \put(105,10.5){\normalsize  \boldmath$\theta_{t2}^*$}

   \put(62.67,14){\vector(0,1){10}}
   \put(65.25,14){\oval(5,5)[bl]}

   \put(56.2,14){\vector(0,1){10}}
   \put(53.72,14){\oval(5,5)[br]}

   \put(47,10.3){\normalsize  \boldmath$\theta_{0t}^*$}
   \put(67,10.5){\normalsize  \boldmath$\theta_{t1}^*$}

   \put(28.35,14.6){\vector(0,1){8.5}}
   \put(24.5,10.5){\normalsize  \boldmath$\theta_0^*$}

   \put(135,16){\normalsize  \boldmath$\theta_{23}^*$}

\end{picture}
\vspace{8mm}
\figcap{ The order parameter $\langle x \rangle = \langle n \rangle /N$, 
         as a function of $\theta$, when $|\Delta|=0.5$ and all other parameters as 
         in \fig{x_og_V_Delta0}. 
         A dotted vertical  line at
         $\theta=\theta_k$, where $\theta_0 = \theta_0^*$, 
         indicates where the $k$:th branch comes 
         into existence.
          The numerical value of $\theta_0^*$  is $\theta_0^*=1.047$. 
         The numerical value of the $\theta_k$, $k\not=0$,
         is the same as in 
         \fig{x_og_V_Delta0} since this critical $\theta$ only depends on $a$.
         The various branches at the extremum of the
         effective potential, i.e. $V_0(\theta)=V_0(\phi ,\theta(\phi))$, 
         are also shown.
         First-order thermal-to-maser phase transitions occur at  
         the pump parameters $\theta_{tk}^*$ 
         and second-order maser-to-thermal transitions occur 
         at $\theta_0^*$, $\theta_{0t}^*$ and $\theta_{kt}^*$. 
         The circle at $\theta = \theta_{23}^*$ in the figure indicates a  maser-to-maser
         transition.
         The numerical values of the critical transition parameters in the
         figure are: 
         $\theta^*_0\approx1.047$, $\theta^*_{0t}\approx5.236$, $\theta^*_{t1}
         \approx6.193$
         $\theta^*_{1t}\approx11.519$, $\theta^*_{t2}\approx11.906$ and  
         $\theta^*_{23}\approx17.425$. 
\label{x_og_V_Delta05} }
\end{figure}

   Critical points where new extrema of $V_0(x)$ 
   appear are determined by $V_0^{\prime \prime}(x)=0$, i.e. 
   non-trivial solutions of  

\be
  \tan \phi=\phi ~~,
\ee

   \noi 
   independent of the physical 
   parameters of the micromaser. This equation 
   has infinitely many positive solutions $\phi=\phi_k$, where
   $k=1,2,...\,$. 
   Corresponding to these solutions we have the critical pump parameters  
   $\theta_k \equiv \theta(\phi_k)$, i.e

\be \label{theta_k}
  \theta_k = \frac{1}{\sqrt{2a-1}} \frac{\phi_k}{|\sin \phi_k|}~~,~~k=1,2,3,...~~, 
\ee

   \noi
   independent of $n_b$ and $\Delta$, for which the $k$'th branch comes 
   into existence. 
   For a given branch $k\geq1$, the $V_0^{\prime \prime}(x(\phi))>0$
   sub-branches and the  $V_0^{\prime \prime}(x(\phi))<0$ sub-branch coincide at 
   this particular critical point $\theta_k$ (see e.g. \fig{x_og_V_Delta0}).
  The critical point $\theta=\theta_k$ also corresponds to $x^{\prime}(\theta_k) = \infty$
   as seen from \eq{der_x_theta}. 

Each sub-branch with $V_0^{\prime \prime}(x(\phi)) > 0$
   of the effective potential $V_0(\theta)$ has one and only one minimum. This can be seen from 
   the expression

\be \label{V_der}
  \frac{d \, V_0(\theta)}{d \theta} = - \frac{2}{\theta(\phi)} [ a+n_b(2a-1) ] \, J(\phi) ~~,
\ee

  \noi where

\be
  J(\phi) \equiv \int_{\phi_0+k\pi}^{\phi} ~ d\chi \,
                 \chi^4  
                 \frac{\sin(2\chi)}
                      { \left [ \chi^2 n_b + a \, \theta(\phi)^2 \sin^2 \chi \right ] 
                        \left [ \chi^2(1+n_b) + b \, \theta(\phi)^2 \sin^2\chi \right ] } ~~, 
\ee

  \noi
  and $V_0(\theta) = V_0(\phi,\theta(\phi))$. 
  The upper integration limit $\phi$ must then be chosen according to \eq{PHI_interv} where we, of course, only consider 
  the $V_0^{\prime \prime}(x(\phi))>0$ sub-branch for $k \geq 1$.

  We can now study the actual intersections of the various branches at a common global minimum
by making use of numerical and analytical methods. The procedure is, for example,
  illustrated in \fig{x_og_V_Delta0} and \fig{x_og_V_Delta05}. We find that
  $V_0(\theta)$ is such that if two branches do intersect at a common global minimum, 
  these branches
  correspond to $k$ and $k+1$, i.e. they are consecutive branches. 
  A given branch can also intersect with $V_0(\theta)=0$, i.e. the thermal phase.
  In addition two consecutive branches can intersect when $V_0(\theta)=0$.
  We then have a triple point.

\vspace{0.5cm}
\bc{
\subsection{The Critical Parameters $\theta_{k k+1}^*$,
$\theta_{tk}^*$ and $\theta_{kt}^*$}
}\ec

   If, for a given $a$, $n_b$ and $\Delta$, we have a 
   transition from the maser 
   branch $k$  to the neighbouring maser branch $k+1$, 
   then there exist a $\phi_k^*$ and  $\phi_{k+1}^*$ such that

\be \label{skj_bet}
   V_0(\phi_k^*) = V_0(\phi_{k+1}^*) ~~ \mbox{and} ~~
   \theta(\phi_k^*) = \theta(\phi_{k+1}^*) \equiv \theta_{k k+1}^* ~~,
\ee

  \noi where

\be \label{phi_STJ}
  V_0(\phi_{k}^*) = -\frac{2}{\theta(\phi_{k}^*)^2}   
                  \int_{\theta(\phi_{k}^*)|\Delta|}^{\phi_{k}^* } d\phi \, \phi \, 
                  \ln[\, w(\phi , \theta(\phi_{k}^*) ) \,] ~~, 
\ee

   \noi
   and where $\phi_k^*$ is in the interval $\phi_k < \phi_k^* < (\pi - \phi_0) + k\pi$.
   The reason why we chose $\phi_k^*>\phi_k$ is that we are looking for solutions of 
   \eq{phi_STJ} for which $V_0^{\prime \prime}(x(\phi))>0$.
   The pump parameter according to \eq{skj_bet} can
   be expressed as

\be \label{a_theta}
   \theta_{k k+1}^* = \frac{1}{\sqrt{2a-1}} \frac{\phi_k^*}{|\sin \phi_k^* |} ~~.
\ee

  \noi 
   Furthermore, for this particular combination of the parameters,
   the order parameter $\langle x \rangle = \langle n \rangle/N$ 
   makes a  discontinous change as can be seen from the \eq{x_phi}, i.e.
   the discontinuity in $\langle x \rangle$ 
   is given $\Delta\langle x \rangle = x(\phi_{k+1}^*) -x( \phi_k^*)$. 
   We therefore have a first-order phase transition.

   If, on the other hand, a given maser branch corresponding to a global 
   minima does not intersect with a neighbouring maser branch, it is the
   intersection with the  thermal branch which determines the  
   phase transition (see e.g. \fig{x_og_V_Delta05}).
   For a given branch $k\geq1$, let $\theta_{tk}^* \equiv \theta(\phi_{tk}^*)$ 
   denote the pump parameter of this thermal-to-maser transition. 
   The value of $\phi_{tk}^*$ is then determined by the solution of

\be \label{tk_bet}
  V_0(\phi_{tk}^*) = 0 ~~,
\ee

  \noi where

\be  \label{V_0_tk}
  V_0(\phi_{tk}^*) = -\frac{2}{\theta(\phi_{tk}^*)^2}   
                  \int_{\theta(\phi_{tk}^*)|\Delta|}^{\phi_{tk}^* } d\phi \, \phi \, 
                  \ln[\, w(\phi , \theta(\phi_{tk}^*) ) \,] ~~. 
\ee                    
  \noi
  We determine $\phi_{tk}^*$ numerically,
  where $\phi_k < \phi_{tk}^* < (\pi - \phi_0) + k\pi$ and $k=1,2,3,...\,$.
  The corresponding pump parameters are  $\theta_{tk}^* \equiv
  \theta(\phi_{tk}^*)$, i.e

\be  \label{theta_tk_STJ}
   \theta_{tk}^* = \frac{1}{\sqrt{2a-1}} \frac{\phi_{tk}^*}{|\sin \phi_{tk}^*|}
   ~~,~~k=1,2,3,...~~. 
\ee

  \noi
  Due to the same reasons as discussed above, such  
  thermal-to-maser transitions correspond
  to first-order phase transitions.

  Furthermore, let $\theta_{kt}^* \equiv \theta(\phi_{kt}^*)$ denote the maser-to-thermal transition 
  for a given branch $k\geq0$ (see e.g. \fig{x_og_V_Delta05}).
  The value of $\phi_{kt}^*$ is determined by one of the solutions of

\be \label{kt_bet}
  V_0(\phi_{kt}^*) = 0 ~~,
\ee

  \noi where

\be  \label{V_0_kt}
  V_0(\phi_{kt}^*) = -\frac{2}{\theta(\phi_{kt}^*)^2}   
                  \int_{\theta(\phi_{kt}^*)|\Delta|}^{\phi_{kt}^* } d\phi \, \phi \, 
                  \ln[\, w(\phi , \theta(\phi_{kt}^*) ) \,] ~~. 
\ee

  \noi We see that

\be
  \phi_{kt}^* = (\pi - \phi_0) + k\pi ~~,
\ee

  \noi 
  for all branches $k\geq0$. When $\Delta\neq0$,  \eq{kt_bet} is trivially fulfilled 
  since the upper integration limit is equal to the lower. If, on the other hand,  
  $\Delta=0$, then \eq{kt_bet} is fulfilled since $\theta(\phi_{kt}^*)=\infty$.
  The pump parameter $\theta_{kt}^* \equiv \theta(\phi_{kt})$ is given by

\be  \label{theta_kt_STJ}
  \theta_{kt}^* = \frac{(k+1)\pi - \arcsin(|\Delta|/\sqrt{2a-1})}{|\Delta|}
   ~~,~~k=0,1,2,...~~.
\ee

   \noi
   This equation is equivalent to

\be   \label{theta_kt_STJ_a}
   a = \frac{1}{2} + \frac{\Delta^2}{2 \sin^2(\theta_{kt}^*\Delta)}~~,~~
   k=0,1,2,...~~,
\ee

   \noi
   which is of the same form as \eq{a_sfa_theta}. 
   Due to the same reasons as discussed above, such a
   maser-to-thermal transition corresponds to a 
   second-order phase transition.

   Equipped with these results, we can now construct a 
   complete phase diagram in e.g. the $a$- and 
   $\theta$-parameter space for a given $n_b$ and $\Delta$.

\begin{figure}[t]
\unitlength=1mm
\begin{picture}(100,80)(0,0)

\includegraphics{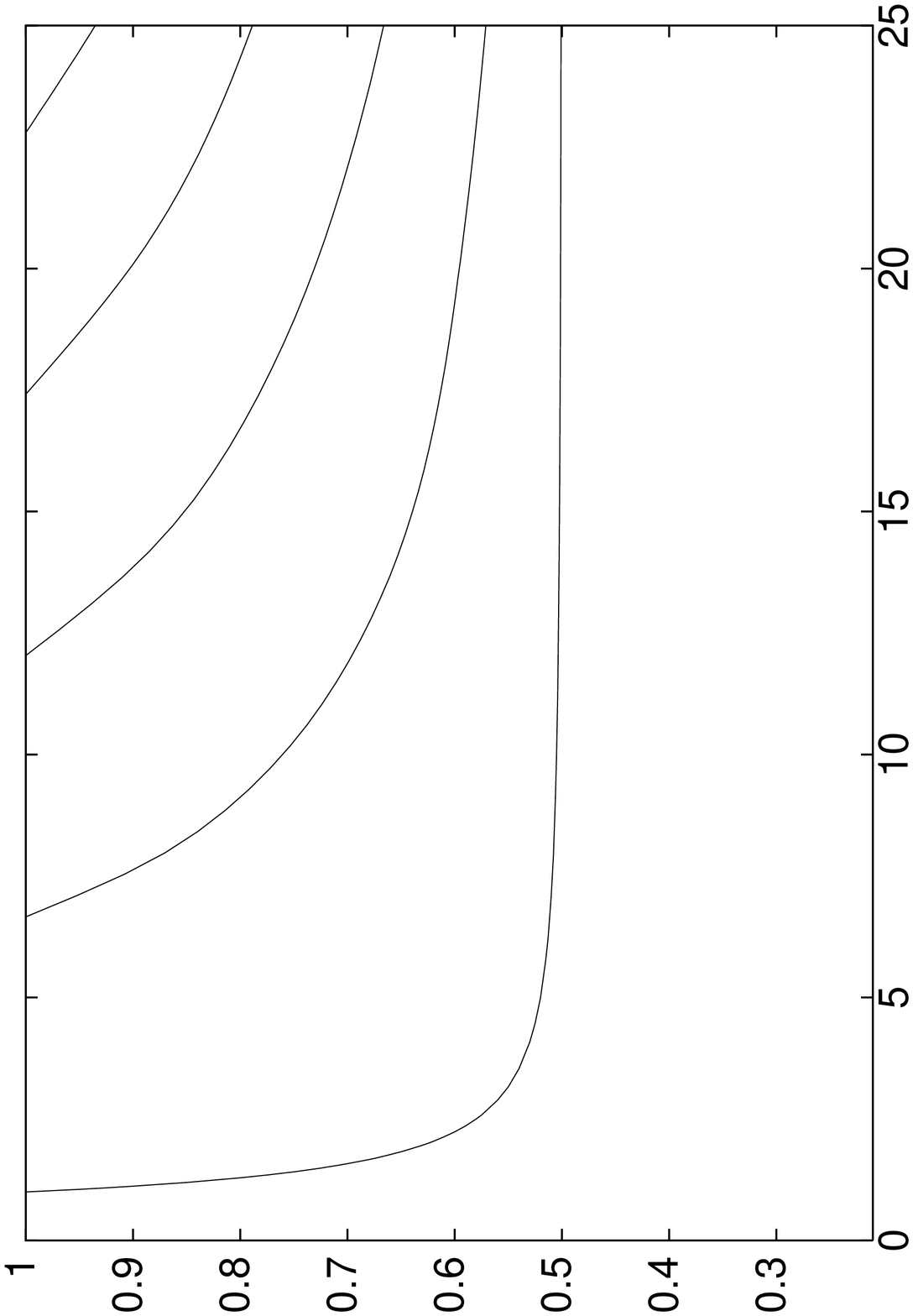}      

   \put(79,-4){\normalsize  \boldmath$\theta$}
   \put(2,55){\normalsize  \boldmath$a$}

   \put(65,25){\normalsize  Thermal phase}

   \put(22,106){\normalsize  \boldmath$\theta_0^*$}
   \put(49,106){\normalsize  \boldmath$\theta_{01}^*$}
   \put(76,106){\normalsize  \boldmath$\theta_{12}^*$}
   \put(102,106){\normalsize \boldmath$\theta_{23}^*$}
   \put(130,106){\normalsize \boldmath$\theta_{34}^*$}

   \put(31,41){\normalsize  \boldmath$\theta_0^*(a)$}
   \put(68,61){\normalsize  \boldmath$\theta_{01}^*(a)$}
   \put(90,73){\normalsize  \boldmath$\theta_{12}^*(a)$}
   \put(110,84){\normalsize \boldmath$\theta_{23}^*(a)$}
   \put(127,93){\normalsize \boldmath$\theta_{34}^*(a)$}

\end{picture}
\vspace{2mm}
\figcap{ The phase diagram for the micromaser system when 
         $n_b=0.15$ and  $\Delta=0$. All the critical lines converge to
         $1/2$ in the large $\theta$ limit. 
         The thermal-to-maser critical line $\theta_0^*(a)$ is 
         determined analytically by \eq{a_sfa_theta_0}.
         The other critical lines $\theta_{k k+1}^*(a)$, which are maser-to-maser 
         transitions, are determined by \eq{a_theta}.
\label{phase_fig_Delta0} }
\end{figure}

\vspace{0.5cm}
\bc{
\subsection{Phase Diagram}
}\ec

   The phase diagram when $n_b=0.15$ and $\Delta=0$ is shown in
   \fig{phase_fig_Delta0}. The first critical line in this figure is 
   given by \eq{a_sfa_theta_0}.
   As already mentioned, this corresponds to a second-order thermal-to-maser 
   phase transition. In the region above this line the mean 
   occupation number $\langle n \rangle$ grows proportionally with the 
   pumping rate $N$. The cavity therefore behaves like  a maser in this regime.
   For values of $a$ and $\theta$ below this line 
   the only global minimum of $V_0(x)$ occurs at $x=0$.
   In this particular region the probability ${\bar p}_n$ is given by the thermal 
   distribution in \eq{p_n_geo}, i.e. the micromaser is in the thermal 
   phase. 
   We remark too that \eq{a_sfa_theta} also determines the 
   radius of converegence of the thermal probability distribution in 
   \eq{p_n_geo}.
 The other critical lines in \fig{phase_fig_Delta0} are determined by 
   \eq{skj_bet} since for $n_b=0.15$ and $\Delta=0$,
   all neighbouring $V_0^{\prime \prime}(x(\phi))>0$
   sub-branches of $V_0(\theta)=V_0(\phi(\theta), \theta)$ intersect for some $a>1/2$,
   see \fig{x_og_V_Delta0}.

\begin{figure}[t]
\unitlength=1mm
\begin{picture}(100,80)(0,0)

\includegraphics{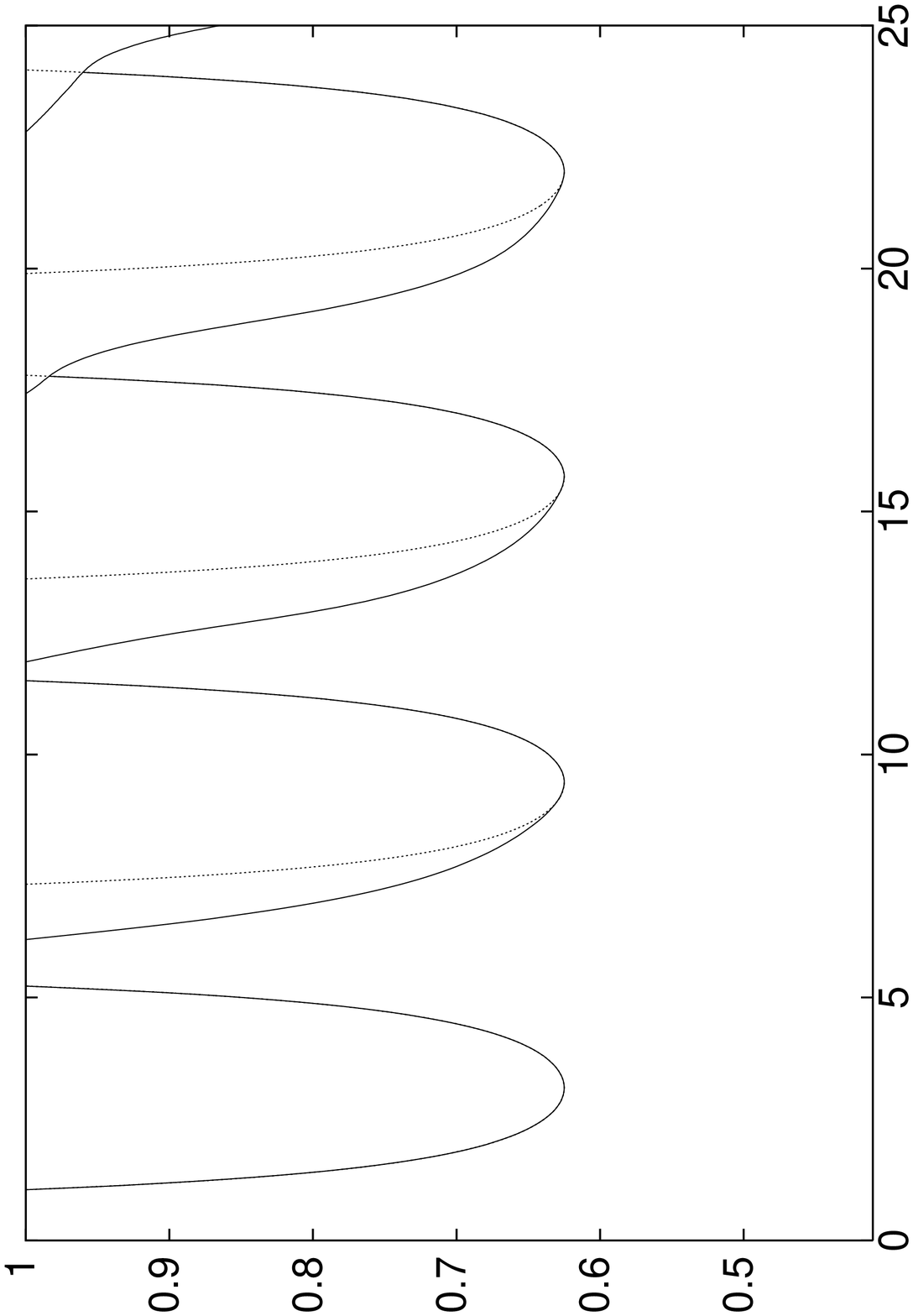}      

   \put(79,-4){\normalsize  \boldmath$\theta$}
   \put(4,54){\normalsize  \boldmath$a$}

   \put(65,19){\normalsize  Thermal phase}

   \put(107.9,99.3){\circle{2.7}}
   \put(139.6,95.8){\circle{2.7}}

   \put(22,106){\normalsize  \boldmath$\theta_0^*$}
   \put(41,106){\normalsize  \boldmath$\theta_{0t}^*$}
   \put(47,106){\normalsize  \boldmath$\theta_{t1}^*$}
   
   \put(75.8,104){\vector(0,-1){1}}
   \put(73.3,104.4){\oval(5,5)[tr]}
   \put(66,106){\normalsize  \boldmath$\theta_{1t}^*$}

   \put(77.8,104){\vector(0,-1){1}}
   \put(80.3,104.4){\oval(5,5)[tl]}
   \put(82,106){\normalsize  \boldmath$\theta_{t2}^*$}

   \put(135.3,35){\vector(0,1){16}}
   \put(132.82,35){\oval(5,5)[br]}
   \put(122.3,42){\vector(0,1){5}}
   \put(119.82,42){\oval(5,5)[br]}
   \put(128.8,41){\line(0,1){5}}

   \put(103.3,35){\vector(0,1){16}}
   \put(100.82,35){\oval(5,5)[br]}
   \put(91.3,42){\vector(0,1){5}}
   \put(88.82,42){\oval(5,5)[br]}
   \put(97,41){\line(0,1){5}}

   \put(71.6,35){\vector(0,1){16}}
   \put(69.12,35){\oval(5,5)[br]}
   \put(59.9,42){\vector(0,1){5}}
   \put(57.42,42){\oval(5,5)[br]}
   \put(65.36,41){\line(0,1){5}}

   \put(33.55,41){\line(0,1){5}}

   \put(119,31){\normalsize  \boldmath$\theta_{3t}^*(a)$}
   \put(107,38){\normalsize  \boldmath$\theta_{t3}^*(a)$}
   \put(76,38){\normalsize  \boldmath$\theta_{t2}^*(a)$}
   \put(88,31){\normalsize \boldmath$\theta_{2t}^*(a)$}
   \put(43.5,38){\normalsize  \boldmath$\theta_{t1}^*(a)$}
   \put(56,31){\normalsize \boldmath$\theta_{1t}^*(a)$}
   
   \put(28.5,55){\normalsize \boldmath$\theta_0^*(a)$}

\end{picture}
\vspace{2mm}
\figcap{ The phase diagram for the micromaser system when 
         $n_b=0.15$ and $|\Delta| = 0.5$. In this diagram we have 
         plotted the range of validity of the thermal distribution
         (dotted lines). When not visible these dotted lines 
         overlap with the solid critical lines. 
         The minima of the critical lines are determined by the 
         condition $\sin^2( \Delta \theta )=1$ and are marked by a short 
         vertical lines. The first critical line $\theta_0^*(a)$
         in this phase diagram is determined analytically by 
         \eq{theta_0_stj}. The critical lines $\theta_{tk}^*(a)$ are 
         given by \eq{theta_tk_STJ} and $\theta_{kt}^*(a)$
         are given by \eq{theta_kt_STJ}.
         Triple points are indicated by circles.
         The numerical values of these triple points are $(a_{23}, \theta_{23})^{triple}=(0.98,17.78)$ 
         and $(a_{34},\theta_{34})^{triple}=(0.96,24.04)$.

\label{phase_fig_Delta05} }
\end{figure}

   When the micromaser is detuned the phase diagram is more complicated.
   \fig{phase_fig_Delta05} shows a typical example of a phase diagram 
   when $\Delta\neq0$.
   As we can see from this figure, the first two maser phases are well separated. 
   The critical value of $\Delta$ for
   such a separation of phases is in general determined by considering phase 
   separation on the line $a=1$. For a given $n_b$, let $\Delta=\Delta_{k k+1}$
   be the corresponding critical value for phase separation. 
   With the definition $\phi_{t0}^*\equiv \phi_0$, where  $\phi_0$ is given in 
   \eq{phi_0}, $\Delta_{k k+1}$ 
   is determined by the transcendental equation

\be 
  \theta(\phi_{kt}^*) = \theta(\phi_{tk}^*) ~~, ~~k=0,1,2,... ~~,   
\ee

  \noi i.e. the solution of

\be
  |\Delta| = \frac{|\sin \phi_{tk}^*|}{\phi_{tk}^*} \left[  \, (k+1)\pi - 
  \arcsin |\Delta|   \, \right ] ~~.  
\ee

   \noi
   Here $\phi_{tk}^*$ is determined numerically according to \eq{tk_bet}.
   For $n_b=0.15$
   the first critical value of detuning is $|\Delta_{01}| \approx 0.408$. 
   When the detuning is larger than $\Delta_{k k+1}$, branch 
   $k$ will separate from branch $k+1$ (see e.g. \fig{phase_fig_Delta05}).
   In passing, we notice that $\Delta_{k k+1}$ 
   is restricted by $\Delta_{k k+1}^2<1$, for all 
   possible branches.

     The first critical line in \fig{phase_fig_Delta05} 
     corresponds to a second-order thermal-to-maser (maser-to-thermal) 
     transition to the left (right) of its minimum.
     The minima of the critical lines are marked by short vertical lines.
     For any other critical line in \fig{phase_fig_Delta05} 
     the transition is a first-order thermal-to-maser 
     (second-order maser-to-thermal) phase transition 
     to the left (right) of its minimum unless it intersects with
     another critical line. 
     In the region between any thermal-to-maser line and the
     neighbouring dotted line, the 
     thermal probability distribution does not correspond to a global
     minimum of $V_0(x)$.
     The phase diagram in \fig{phase_fig_Delta05} also contains triple points. 
     These points are marked by circles. They are mathematically determined by

\be \label{trippel_bet} 
 V_0(\phi_{tk}^*) = V_0(\phi_{kt}^*) = 0 ~~ \mbox{and} ~~  
 \theta(\phi_{tk}^*) = \theta(\phi_{kt}^*)~~.
\ee

\newpage
\vspace{1cm}
%
\bc{
\section{THE ORDER PARAMETER} 
\label{order_param_KAP}
}\ec

    When the system is in a maser phase
    the probability distribution is given by \eq{p_j_gauss}.
    $\langle x \rangle = \langle n\rangle /N$ is then, of course, trivial to compute.
    For a given $a$, $\Delta$ and $\theta$    $\langle x \rangle$ is simply given by

\be \label{x_maser}
   \langle x \rangle = x_j~~,
\ee

   \noi
   where $x_j$ is the value of $x$ where $V_0(x)$ reaches its global minimum,
   i.e. $x_j$ is one of the saddle-points as determined by \eq{trans}.
   In the maser regime, the photon number 
   behaviour (see e.g. \fig{x_og_V_Delta0}) is well-established
   in the 
   literature (see e.g. Refs.\cite{Walther88}).

   For values of $a$ and $\theta$ corresponding to the first critical line 
   \eq{theta_0_stj}, the order parameter $\langle x \rangle$ is zero. 
   Mathematically this follows trivially 
   by substituting $\phi_0$ in \eq{phi_0} into \eq{x_phi}
   and using \eq{x_maser}.
   Below this critical line, the 
   micromaser is in the thermal phase and the order parameter is then given by

\be \label{x_termisk}
  \langle x \rangle = \frac{1}{N} \frac{n_b + a \, \theta_{eff}^2}{1+(1-2a) 
  \, \theta_{eff}^2}~~,
\ee

   \noi
   which is zero in the large $N$ limit.
   We realize from \eq{x_termisk} 
   that the photon number $\langle n \rangle$ is in general
   a periodic function of $\theta$ when the system is 
   detuned. The period of $\langle n \rangle$ is then $|\Delta|\theta$.
   It reaches its maximum $(n_b + a/\Delta^2)/(1+(1-2a)/\Delta^2)$ for 
   $\Delta\theta = (n+1/2)\pi$, where $n=0,1,...$\,. The minimum of $\langle n \rangle$
   is $n_b$ and occurs for $|\Delta|\theta = n\pi$. 
   In passing, we note that the maximum value $(n_b + a/\Delta^2)/(1 + (1-2a)/\Delta^2)$ 
   is reduced to $n_b$ in thermal equilibrium, for any $n_b$, $\Delta$  and $\theta$.
   We also note that when $n_b=0$ the cavity is dark, i.e.   $\langle x \rangle =0$
   for every integer multiple of 
   $\theta = \pi/|\Delta|$.

\begin{figure}[t]
\unitlength=1mm
\begin{picture}(100,80)(0,0)

\includegraphics{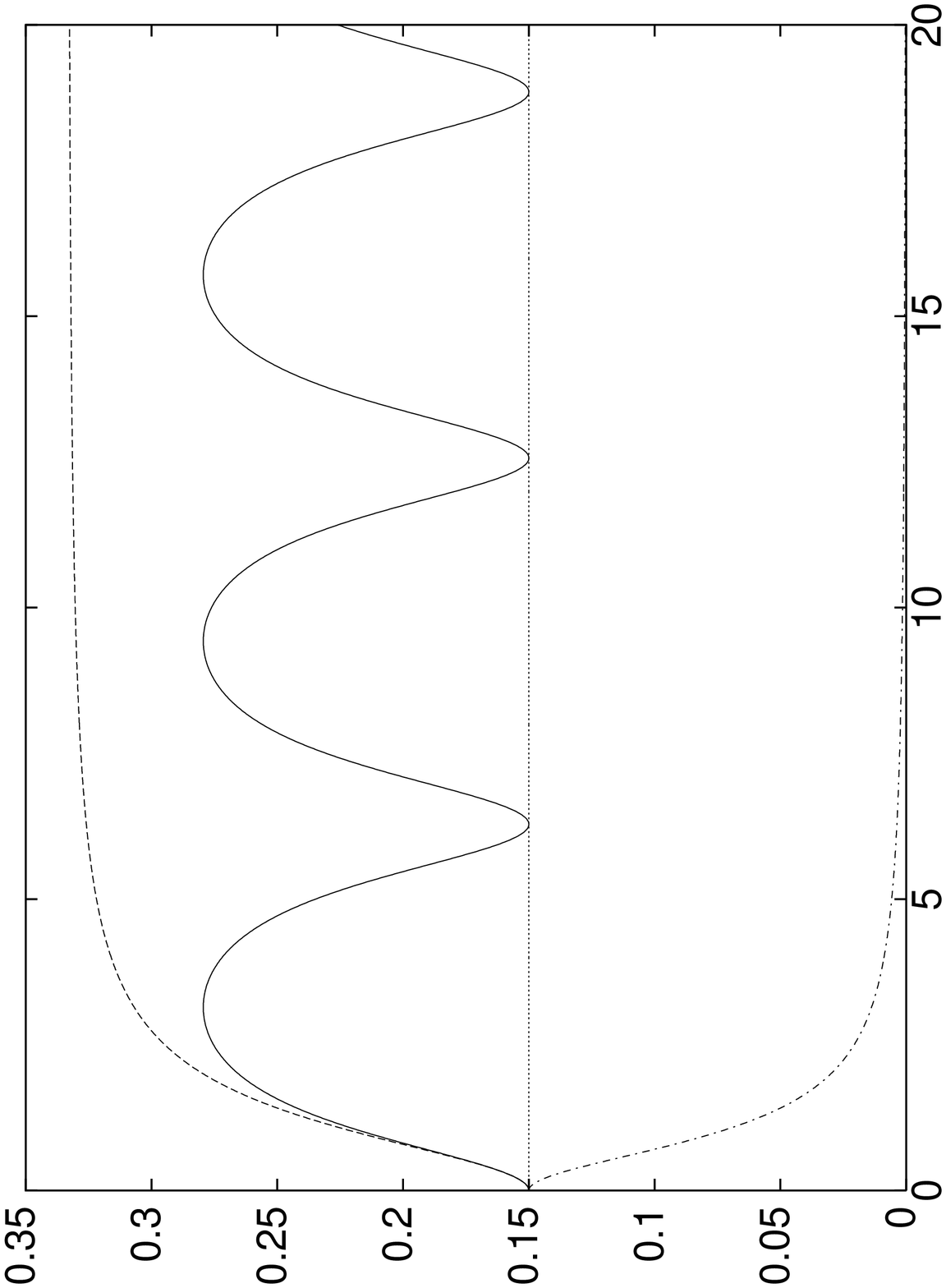}      

   \put(79,-4){\normalsize  \boldmath$\theta$}
   \put(-5,55){\normalsize  \boldmath$\langle n \rangle$}

\end{picture}
\vspace{2mm}
\figcap{ 
         This figure shows the different behaviours of \eq{x_termisk}. 
         $n_b=0.15$ for all curves.
         The straight line corresponds to $a=n_b/(1+2 \, n_b) = 3/26$,
         i.e. thermal equilibrium. The solid curve corresponds to $a=0.2$ and 
         $|\Delta|=0.5$. The fat dotted curve corresponds to $a=0.2$ and $\Delta=0$.
         The dashed-dotted curve corresponds to $a=0$ and $\Delta=0$. 
\label{n_fig} }
\end{figure}

   When the system is not detuned, \eq{x_termisk} is not 
   oscillating and it then reduces to

\be \label{x_termisk_Delta0}
   \langle n \rangle = \frac{n_b + a \, \theta^2}{1 + (1-2a) \, \theta^2}~~.
\ee

   \noi
   In the large $\theta$ limit, 
   \eq{x_termisk_Delta0} converges to $a/(1-2a)$ (see \fig{n_fig}).

\begin{figure}[t]
\unitlength=1mm
\begin{picture}(100,80)(0,0)

\includegraphics{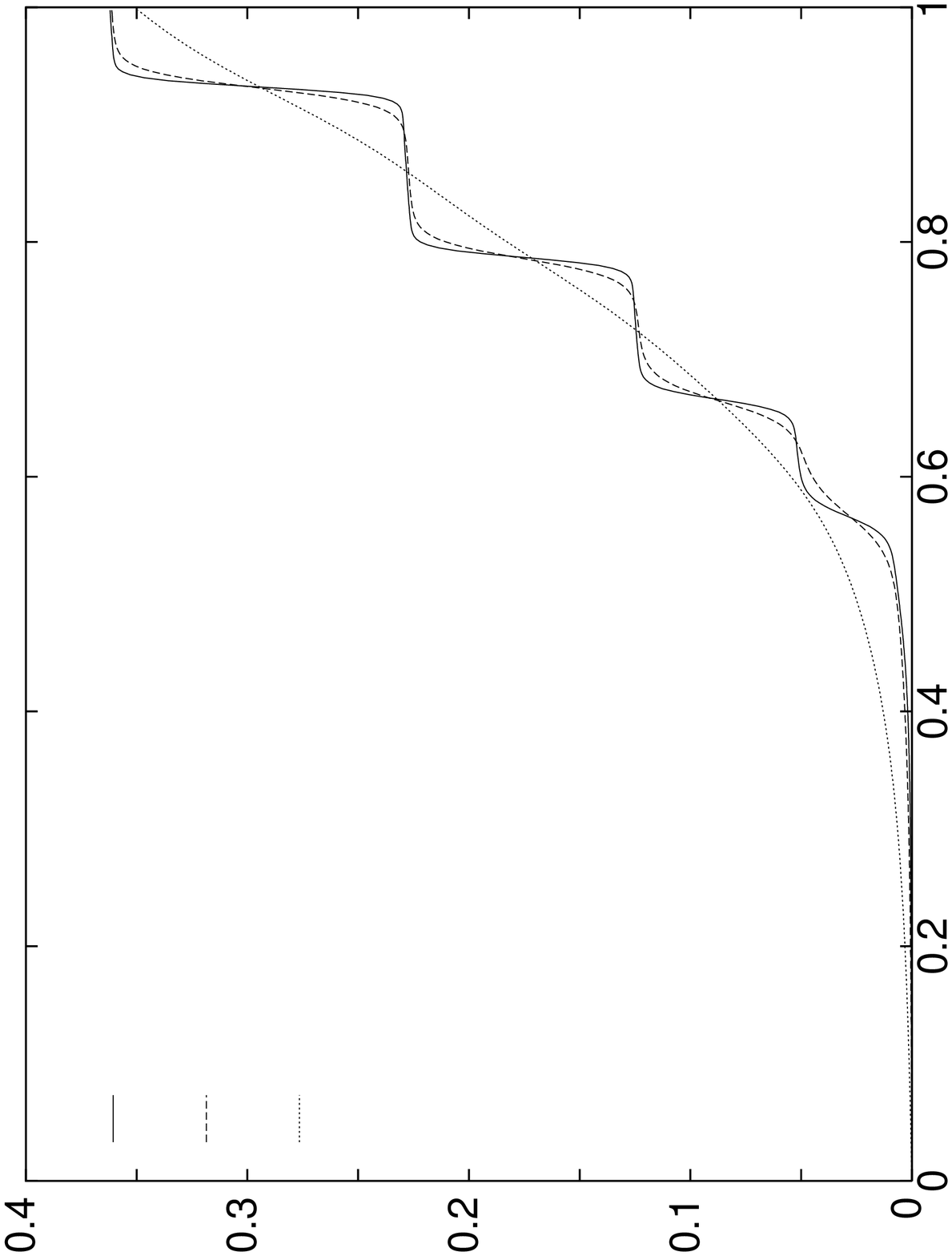}      

   \put(79,-4){\normalsize    \boldmath$a$}
   \put(35,50){\normalsize \boldmath$\langle x \rangle = \langle n \rangle /N$}

   \put(31,93){\normalsize  \boldmath $N=1000$}
   \put(31,83){\normalsize  \boldmath $N=500$}
   \put(31,73){\normalsize  \boldmath $N=100$}

\end{picture}
\vspace{2mm}
\figcap{ The order parameter $\langle x \rangle = \langle n \rangle /N$ as a 
         function of $a$ when $n_b=0.15$, $\Delta=0$, $\theta=25$ and 
         $N=100,500,1000$.  
         Each step in $\langle x \rangle$ corresponds to a point on one of the 
         transition curves in \fig{phase_fig_Delta0}.
\label{ordr_param_fig} }
\end{figure}

   The order parameter   $\langle x \rangle$ can also be studied as a function 
   of the probability $a$ for the pump atoms to be in an excited state. 
   Varying $a$ over all allowed values keeping other physical parameters fixed,
   the order parameter $\langle x \rangle$ makes a discontinuous change 
   for every combination of the physical parameters corresponding to
   a critical line. 
   Hence, $\langle x \rangle$ as a function of $a$ exhibits a plateau-like behaviour 
   when $N$ is sufficiently large. Such a behaviour is illustrated in \fig{ordr_param_fig}.
   In this figure we observe a rather slow $N$ convergence to this plateau-like behaviour.

\vspace{0.5cm}
%
\bc{
\subsection{Twinkling and Detuning}
}\ec

  Above we have observed that the mean value $\langle n \rangle$ may oscillate 
  as a function of $\theta$
  when the micromaser is in the thermal phase.
  We say that the system is then in a twinkling  mode.
    This twinkling behaviour has a close resemblance to
  the  observed atomic inversion revivals \cite{RWK_87,BSMDHRH_96}.

   A somewhat different twinkling behaviour of the micromaser system
   occurs when the thermal phase intersects 
   with a maser phase. This feature is illustrated in 
   \fig{phase_fig_Delta05} with $|\Delta|=0.5$ and $n_b=0.15$. 
   For a given $a>1/2 + \Delta^2/2$ we then see that the system has 
   repeated thermal-to-maser and maser-to-thermal transitions as $\theta$ increases. 
   This twinkling behaviour is not strictly
   periodic in $\theta $. The
 winkling phenomena will now, however,   
   be more pronounced since, for large $N$, the maser will be
   ``dark'' in the thermal phase, but $\langle n \rangle$ is large in the maser phase since 
   $\langle n \rangle$ is proportional to $N$ in this region.

\vspace{1cm}
\bc{
\section{THE CORRELATION LENGTH}
\label{corr_KAP} 
}\ec

   In this section we study long-time correlations in the large $N$ limit.
   This notion of a correlation length for the micromaser system
   was first introduced in Refs. \cite{ElmforsLS95}.
   These correlations have a surprisingly rich structure and reflect global 
   properties of the stationary photon distribution. 
   The probability ${\cal P}(s)$ of finding an atom in a state 
   $s=\pm$ after the interaction with the cavity, where $+$ represents 
   the excited state and $-$ represents the ground state, can  
   be expressed in the following matrix form \cite{ElmforsLS95}:

\be
  {\cal P}(s) = {\bar u}^{0^T} M(s) {\bar p} ~~,
\ee

   \noi
   such that ${\cal P}(+)\,+\,{\cal P}(-) = 1$.
   The elements of the vector ${\bar p}$ are given 
  the equilibrium distribution in \eq{p_n_eksakt},
   and the matrix $M(s)$ is given by Eqs.(\ref{M_plus}) and (\ref{M_minus}).
   The quantity ${\bar u}^0$ is a vector with all entries equal to $1$, ${\bar u}_n^0=1$.
   Furthermore, the joint probability for observing
   two atoms in the states $s_1$ and $s_2$ with a time-delay $t$ between 
   them, is given by \cite{ElmforsLS95}

\bea
  {\cal P}(s_1,s_2,t) &=& {\bar u}^{0^T} S(s_2) \,  
  e^{-\gamma L t} \, S(s_1)~ {\bar p} \nonumber \\ 
  &=&
  {\bar u}^{0^T} M(s_2) \,  
  e^{-\gamma L t} \, S(s_1)~ {\bar p} ~~,
\eea

  \noi where $L$ is given by \eq{L_matise} and where

\be
   S(s) = (1+L_C/N)^{-1}M(s) ~~.
\ee

   \noi
   We observe that ${\cal P}(+,-,t) = {\cal P}(-,+,t)$ \cite{ElmforsLS95},
   which means that the cavity photons and the outgoing atoms are not 
   quantum-mechanically entangled. Statistical correlations do, however, exist.

  A properly normalised correlation function is then defined by

\be \label{gamma_A}
  \gamma^A(t) = \frac{ \langle ss \rangle _t - \langle s \rangle ^2}
                     {  1 - \langle s \rangle ^2 }~~,
\ee

   \noi
   where $\langle ss \rangle_t = \sum _{s_1,s_2} s_1 s_2 {\cal P}(s_1,s_2,t)$
   and $\langle s \rangle = \sum _s s \, {\cal P}(s)$. This correlation
   function satisfies $-1 \leq \gamma^A(t) \leq 1$. At large times 
   $t\rightarrow \infty$, we then define the atomic beam correlation length 
   $\xi_A$ by \cite{ElmforsLS95}

\be \label{gamma_A_exp}
   \gamma_A(t) \simeq e^{-t/\xi_A}~~.
\ee

\begin{figure}[t]
\unitlength=1mm
\begin{picture}(100,80)(0,0)

\includegraphics{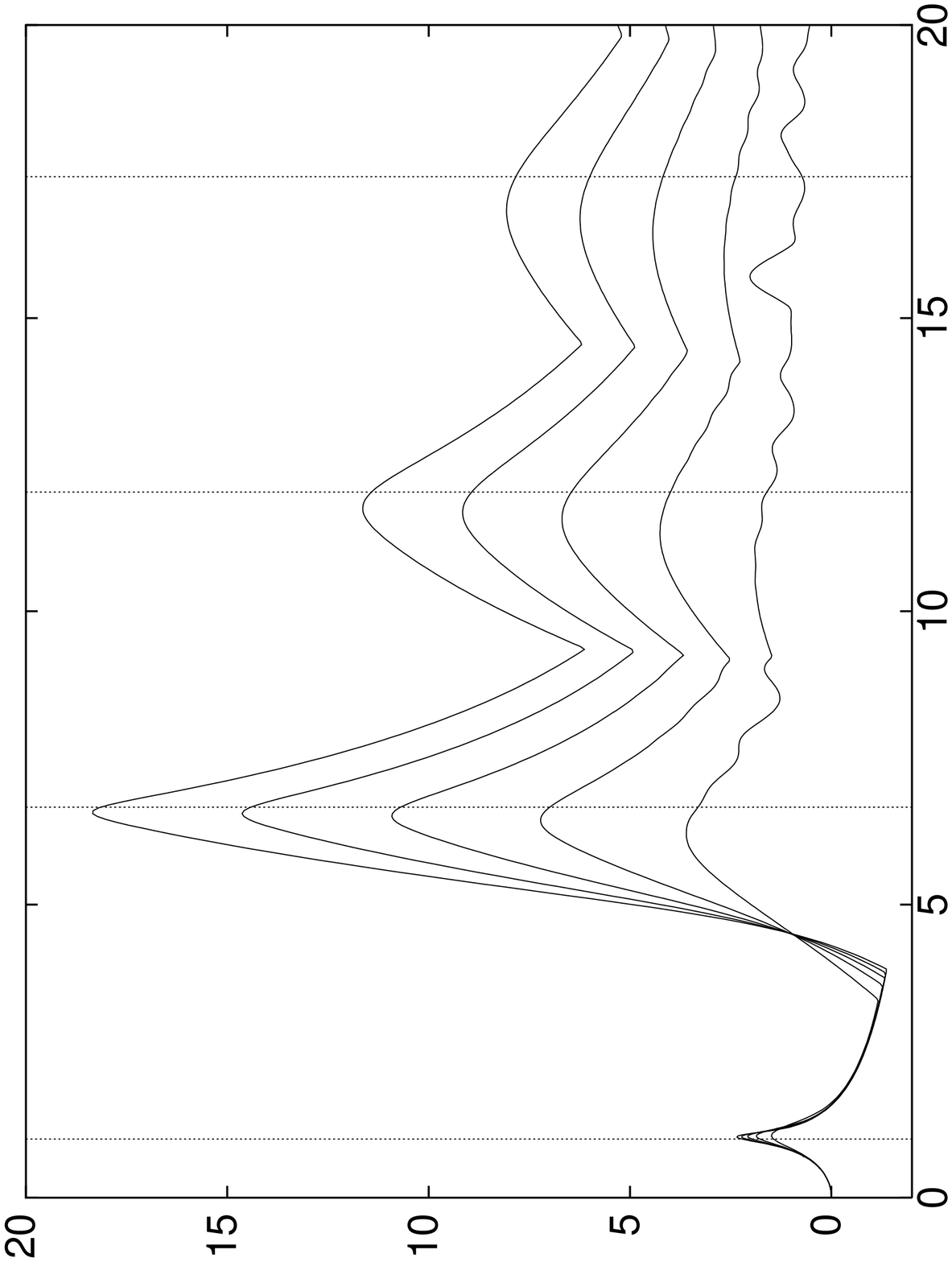}

   \put(79,-4){\normalsize  \boldmath$\theta$}
   \put(-5,55.5){\normalsize  \boldmath$\log(\gamma \xi)$}

   \put(25,108){\normalsize  \boldmath$\theta_0^*$}
   \put(57,108){\normalsize  \boldmath$\theta_{01}^*$}
   \put(90.1,108){\normalsize  \boldmath$\theta_{12}^*$}   
   \put(122.4,108){\normalsize  \boldmath$\theta_{23}^*$}

\end{picture}
\vspace{2mm}
\figcap{ The logarithm of the correlation length  $\gamma \xi$ as a function of 
         $\theta$ for various values of $N=25,50,...,125$ when $n_b=0.15$, 
         $\Delta=0$ and $a=1$. The numerical values of the vertical lines 
         are as in \fig{x_og_V_Delta0}.
\label{corr_fig} }
\end{figure}

   \noi
   The lowest eigenvalue $\lambda = 0$ of $L$ 
   determines the stationary equilibrium solution ${\bar p}$. 
   The next non-zero eigenvalue $\lambda_{nz}$ of $L$,
   on the other hand, determines the typical time scales for the approach 
   to the stationary situation. This eigenvalue can be determined numerically.
   The relation between $\lambda_{nz}$ and the atomic beam correlation length
   is  $1/\xi_A = \gamma \lambda_{nz}$.
   For photons we define a similar correlation length $\xi_{C}$. It
   follows that the correlation lengths are identical, i.e. 
   $\xi_A=\xi_C\equiv \xi$ \cite{ElmforsLS95}.

\begin{figure}[t]
\unitlength=1mm
\begin{picture}(100,80)(0,0)

\includegraphics{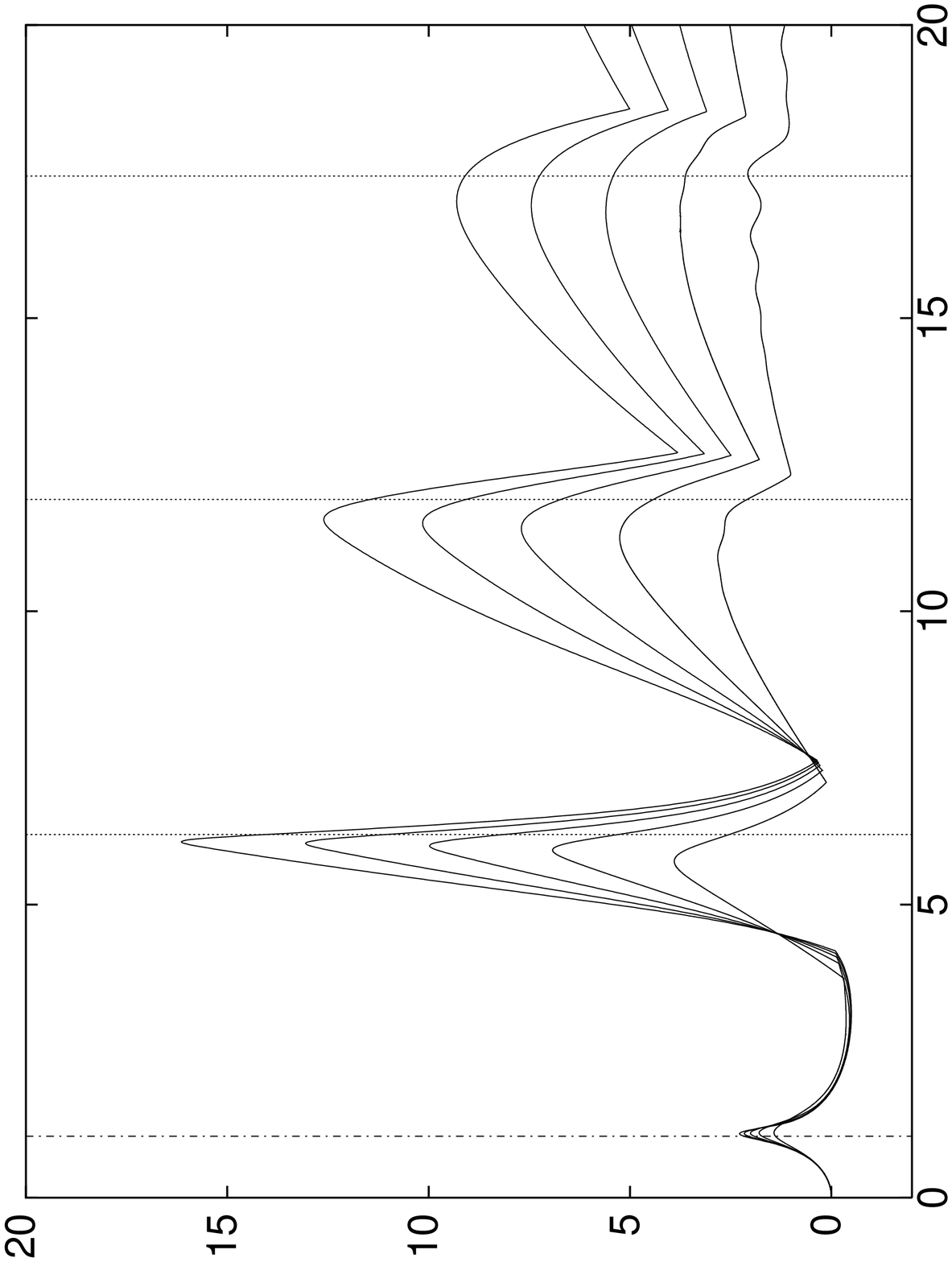}      

   \put(79,-4){\normalsize  \boldmath$\theta$}
   \put(-5,55.5){\normalsize  \boldmath$\log(\gamma \xi)$}

   \put(25,108){\normalsize  \boldmath$\theta_0^*$}
   \put(57,108){\normalsize  \boldmath$\theta_{t1}^*$}
   \put(90.1,108){\normalsize  \boldmath$\theta_{t2}^*$}   
   \put(123.8,108){\normalsize  \boldmath$\theta_{23}^*$}   
 
\end{picture}
\vspace{2mm}
\figcap{ The logarithm of the correlation length  $\gamma \xi$ as a function of 
         $\theta$ for various values of $N=25,50,...,125$ when $n_b=0.15$, 
         $|\Delta|=0.5$ and $a=1$. The numerical values of the vertical lines 
         are as in \fig{x_og_V_Delta05}.
\label{corr_Delta05_fig} }
\end{figure}

\begin{figure}[t]
\unitlength=1mm
\begin{picture}(100,80)(0,0)

\includegraphics{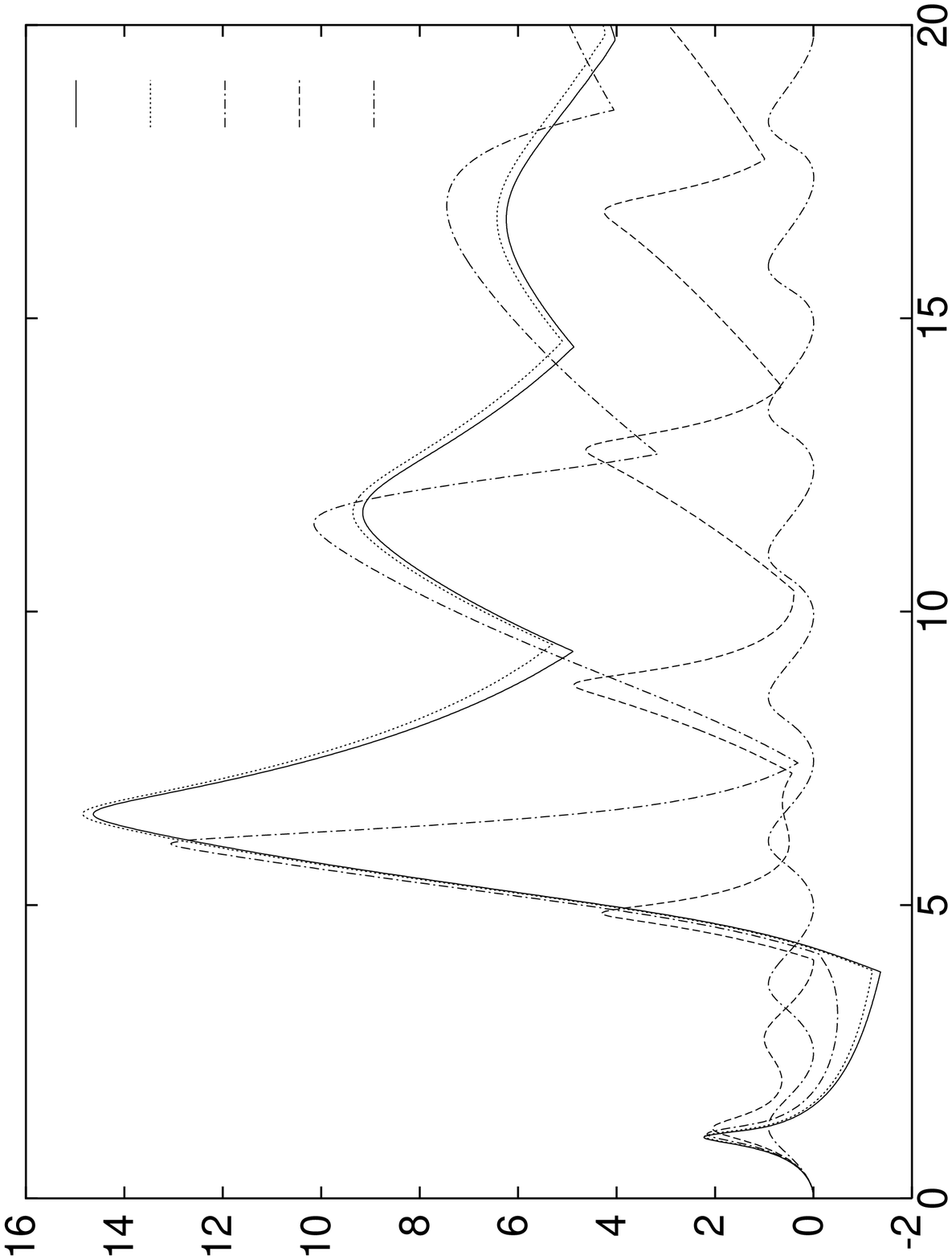}      

   \put(79,-4){\normalsize  \boldmath$\theta$}
   \put(-5,55.5){\normalsize  \boldmath$\log(\gamma \xi)$}
 
   \put(105,97){\normalsize  \boldmath$\Delta = 0$} 
   \put(105,89){\normalsize  \boldmath$|\Delta| = 0.25$} 
   \put(105,81){\normalsize  \boldmath$|\Delta| = 0.5$} 
   \put(105,73){\normalsize  \boldmath$|\Delta| = 0.75$} 
   \put(105,65){\normalsize  \boldmath$|\Delta| = 1.25$} 

\end{picture}
\vspace{2mm}
\figcap{ The logarithm of the correlation length  $\gamma \xi$ as a function of 
         $\theta$ for various values of $|\Delta|=0, \, 0.25, \, 0.50, \, 0.75$ and 
         $1.25$ when $n_b=0.15$, $a=1$ and $N=100$. 
\label{corr_Delta_fig} }
\end{figure}

   The correlation length $\gamma \xi$ is shown in \fig{corr_fig}
   ($\Delta=0$) and \fig{corr_Delta05_fig} ($|\Delta|=0.5$) for various values
   of $N$. Furthermore, \fig{corr_Delta_fig} shows $\gamma \xi$ for various 
   values of the detuning.
   When the detuning is sufficiently small, we observe 
   from these figures that
   the correlation length exhibits large peaks for different values of $\theta$.
   In the large $N$ limit, numerical studies reveal  that these large peaks occur at the 
   transition parameters $\theta_0^*$, $\theta^*_{k k+1}$ and/or $\theta^*_{tk}$,
   depending on the values of the given $n_b$ and $\Delta$ (see e.g. 
   \fig{corr_fig} and \fig{corr_Delta05_fig}).  
   We will discuss the behaviour of these peaks in more detail in Section
   \ref{tunnel_KAP}.
   When the detuning is sufficiently large, i.e. $\Delta^2 > 2a-1$, the correlation 
   becomes smaller and behaves in a strictly periodic manner 
   as a function of the pump parameter $\theta$ (see e.g. \fig{corr_Delta_fig}).

   In order to arrive at a better understanding of the behaviour of the
   correlation length, we will derive various approximative expressions in 
   the following subsections. 
      We will use three different methods in order to derive such
   expressions for $\xi$ and we will compare them with the exact numerical results.

\vspace{1.0cm}
\bc{
\subsection{The Eigenvalue Value Method}
}\ec

   By making use of the orthonormality conditions for
   the left ($u_n$) and right ($p_n$) eigenvectors corresponding to 
   the eigenvalue problem as defined by \eq{master_eq}, i.e. $u^{T}L = \lambda u^T$
   and $Lp=\lambda p$, it is shown in 
   \cite{ElmforsLS95} that an eigenvalue satifies the equation

\be
  \lambda = \sum_{n=0}^{\infty} \, p_n B_n( u_n - u_{n-1})^2~~,
\ee

   \noi  where 

\be
   B_n = (n_b+1)n + N b q_n~~.
\ee

   \noi
   and $p_n ={\bar p}_n^0 u_n$.
   The left eigenfunction $u_n$ corresponding to the first non-zero eigenvalue 
   $\lambda_{nz}$ has one node. It is natural to assume that
   this node occurs at $\langle n \rangle$. 
 This is so since in the neighbourhood of $\langle n \rangle$ we can then use the Ansatz 

\be
\label{ansatz}
   u_n = \frac{n - \langle n \rangle}{\sigma_n}~~,
\ee

  \noi
  where $\sigma^2_n = \langle n^2 \rangle - \langle n \rangle^2$ as usual. 
  The average values are assumed to be taken over the equilibrium distribution ${\bar p}_n$.
  It then follows that $u\cdot {\bar p}= \langle u_n \rangle =0 $ 
  and $u\cdot p = \langle u_{n}^{2}\rangle =1$. 
  By the Ansatz Eq.(\ref{ansatz}) it now follows that

\be \label{corr_lambda}
  \gamma \xi \simeq \gamma \xi_E(\theta) = \frac {\sigma_n^2}{(n_b+1) \langle n \rangle + N b \, 
  \langle q_n \rangle}~~,
\ee

   \noi
   since $\gamma \xi = 1/\lambda_{nz}$.    
   \eq{corr_lambda} can be simplified in the large $N$ limit. In this case
   we have $\langle q_n \rangle \approx \theta_{eff}^2 \langle n \rangle /N$. 
   If, in addition, the micromaser is in the thermal phase, then \eq{corr_lambda}  
   reduces to

\be \label{corr_termisk_master}
  \gamma \xi_E(\theta) = \frac {1}{1 - (2a-1) \theta_{eff}^2}~~,
\ee

   \noi
   which explains the periodic behaviour of the correlation as
   mentioned in the beginning of this section.
   We also observe that $\gamma \xi_E = 1$ when $a=1/2$, independent of any 
   of the other physical parameters at hand.

\vspace{0.5cm}
\bc{
\subsection{The Master Equation Method}
}\ec

   Let us now derive an approximative expression for the correlation length
   by making use of  another method.
   For a general right eigenvector of the matrix $L$ we define $p(x)=N p_n$ 
   and write it as $p(x)={\bar p}(x) u(x)$ with the left 
	eigenvector $u(x)=u_n$.
   It is then shown in \cite{ElmforsLS95} that

\be \label{DL}
  \lambda \, u(x) = \left[\, x - (2a-1)\,q(x) \, \right] \frac{d u(x)}{dx}  - 
               \frac{1}{N} 
               \frac{d}{dx} \left[ (\, n_b x + a q(x) \,) \frac{d u(x)}{dx} 
               \right]~~,
\ee

   \noi
   in the large $N$ limit.
   Here we   again make use of the fact that the eigenfunction corresponding 
    to $\lambda_{nz}$ has 
   only one node, say at $x=x_0$. The function $u(x)$ is therefore 
   given by $u(x) \approx x-x_0$ in the neighbourhood of $x_0$. Furthermore, 
   when $\theta \sqrt{x(\theta)+\Delta^2} \ll 1$, we can expand $q(x)$
   around $x=0$, in which case \eq{DL} reduces to

\be \label{lambda_x0}
   \lambda_{nz} (x-x_0) = \alpha(\theta) \, x + \beta(\theta) \, x^2 - \frac{\gamma(\theta)}{N}~~,
\ee

   \noi
   where

\be  
  \alpha(\theta) = 1 - (2a-1) \, \theta_{eff}^2~~,  
\ee

\be
   \beta(\theta) = (2a-1)
                   \left[ \, \frac{\sin^2(\theta \Delta)}{\Delta^4} - 
                           \frac{\theta \sin(\theta|\Delta|)  
                           \cos(\theta|\Delta|)}{\Delta^3} \, \right]~~,
\ee

\be \label{gamma_theta}
   \gamma(\theta) = n_b + a \, \theta_{eff}^2~~.     
\ee

   \noi
   Notice that \eq{gamma_theta} is only valid when we are not too close to 
   a critical line.
   For small perturbations around $x_0$, i.e. $x = x_0 + \delta x$,
   the eigenvalue $\lambda_{nz}$ is then

\be \label{lambda_alpha}
  \lambda_{nz} = \alpha(\theta) + 2 \beta(\theta) \, x_0~~,
\ee

   \noi
   where $x_0$ is easily solved by the aid of this equation and \eq{lambda_x0}:

\be   \label{x_0}
  x_0(\theta) = - \frac{\alpha(\theta)}{2 \, \beta(\theta)} 
                + \sqrt{ \left(\frac{\alpha(\theta)}{2 \, \beta(\theta)} \right)^2 + 
                \frac{\gamma(\theta)}{\beta(\theta) \, N} }~~.
\ee

   \noi
   Since $1/\xi = \gamma \lambda_{nz}$, we get

\be \label{corr_approx_lambda}
  \gamma \xi \simeq \gamma \xi_M(\theta)  = \frac{1}{\sqrt{\alpha^2(\theta) + 
              {\displaystyle \frac{  4 \, \beta(\theta) \, \gamma(\theta)}{N}}}} ~~.
\ee

   \noi
   In the large $N$ limit, where $x_0=0$ (see \eq{x_0}), \eq{corr_approx_lambda} 
   is reduced to $\gamma \xi_M(\theta) = 1/\alpha(\theta)$, i.e.

\be \label{corr_lambda_termisk}
  \gamma \xi_M(\theta) = \frac{1}{1 - (2a-1)\theta_{eff}^2}~~,
\ee

   \noi
   in agreement with \eq{corr_termisk_master}.

\begin{figure}[t]
\unitlength=1mm
\begin{picture}(100,80)(0,0)

  \includegraphics{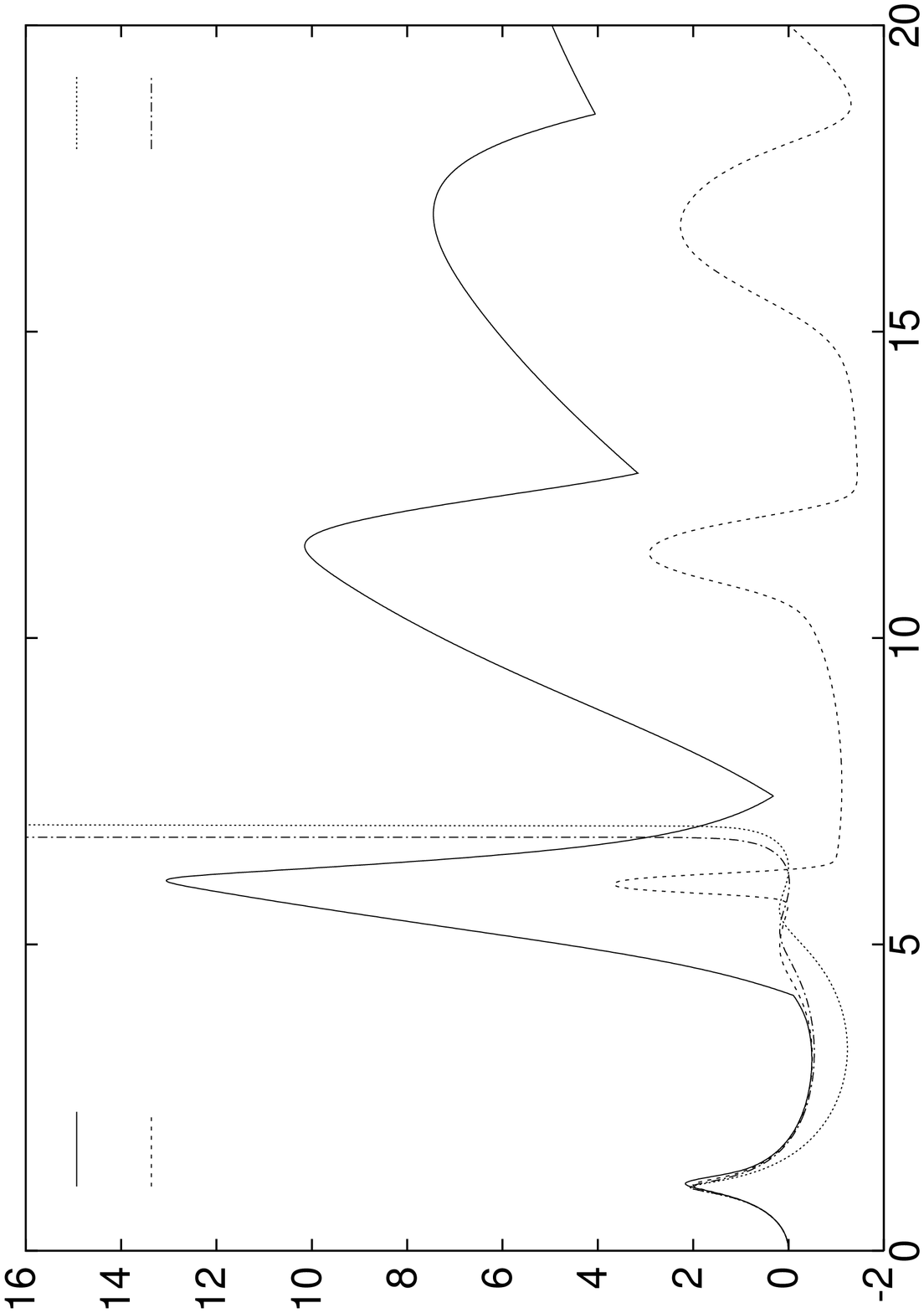}      


   \put(79,-4){\normalsize  \boldmath$\theta$}
   \put(-6,55.5){\normalsize  \boldmath$\log(\gamma \xi)$}

   \put(32,95){\normalsize Exact}
   \put(31.6,87.5){\normalsize \boldmath$\log(\gamma \xi_E)$ }

   \put(105,95){\normalsize  \boldmath$\log(\gamma \xi_M)$}
   \put(105,87.5){\normalsize  \boldmath$\log(\gamma \xi_{MF})$ }

\end{picture}
\vspace{2mm}
\figcap{ Comparison of $\log( \gamma \xi )$ as a function of $\theta$ for various 
         analytical expressions of the correlation. The parameters for
         all lines are $a=1$, $|\Delta|=0.5$, $n_b=0.15$ and $N=100$.
         The solid curve is the exact correlation length.
\label{corr_a1_Delta05_N100} }
\end{figure}

\begin{figure}[t]
\unitlength=1mm
\begin{picture}(100,80)(0,0)

  \includegraphics{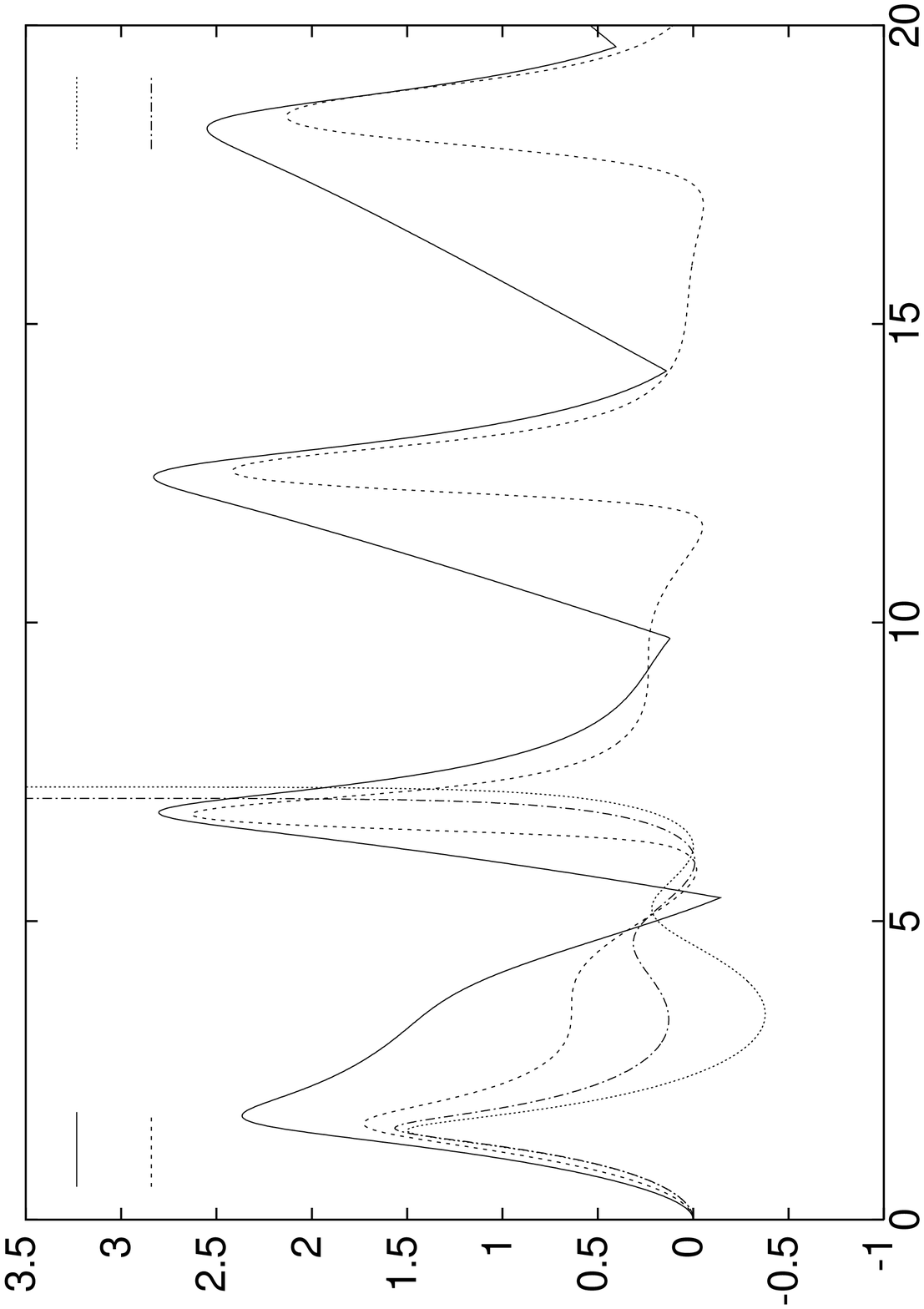}

   \put(79,-4){\normalsize  \boldmath$\theta$}
   \put(-6,55.5){\normalsize  \boldmath$\log(\gamma \xi)$}

   \put(32,95){\normalsize Exact}
   \put(31.6,87.5){\normalsize \boldmath$\log(\gamma \xi_E)$ }

   \put(105,95){\normalsize  \boldmath$\log(\gamma \xi_M)$}
   \put(105,87.5){\normalsize  \boldmath$\log(\gamma \xi_{MF})$ }
\end{picture}
\vspace{2mm}
\figcap{  Comparison of $\log( \gamma \xi )$ as a function of $\theta$ for various 
          analytical expressions derived in the text. The parameters for
          all curves are $a=0.75$, $|\Delta|=0.5$, $n_b=0.15$ and $N=100$.
          The solid curve is the exact correlation length.
         
\label{corr_a075_Delta05_N100} }
\end{figure}

   The correlation length $\xi_M(\theta)$ exhibits 
its first peak at $\theta = \theta_0^*$ (see e.g.
   \fig{corr_a1_Delta05_N100} and  \fig{corr_a075_Delta05_N100}). The
   correlation length \eq{corr_approx_lambda} at this particular peak is

\be \label{corr_chi}
  \gamma \xi_M(\theta_0^*) = \frac{2a-1}{2} \sqrt{ \frac{N}{a + n_b(2a-1)}} 
               ~\chi(\theta_{0}^*|\Delta|)~~,
\ee

   \noi
   where

\be \label{chi}
  \chi(x) = \sqrt{\frac{\sin^2 x} {1- x \cot x  }} ~~.
\ee

   \noi
   In \eq{corr_chi} the probability $a$ can only be chosen in 
   the interval $1/2 + \Delta^2/2 \leq a \leq 1$. 
   This equation shows that the correlation $\theta = \theta_0^*$
   grows as $\sqrt{N}$.
   Again we notice the factor $a + n_b(2a-1)$.
   Also note that \eq{corr_chi} vanishes when $a=1/2$.
   Since by assumption $\theta|\Delta|\ll1$, then \eq{chi} is
   $\chi(\theta |\Delta|) \approx \sqrt{3} \, [ \, 1 - (\theta \Delta)^2/5 \, ]$.
   The correlation length in \eq{corr_chi} is then in agreement with the results of 
   Ref.\cite{ElmforsLS95} for $\Delta =0$.

\begin{figure}[t]
\unitlength=1mm
\begin{picture}(100,80)(0,0)

  \includegraphics{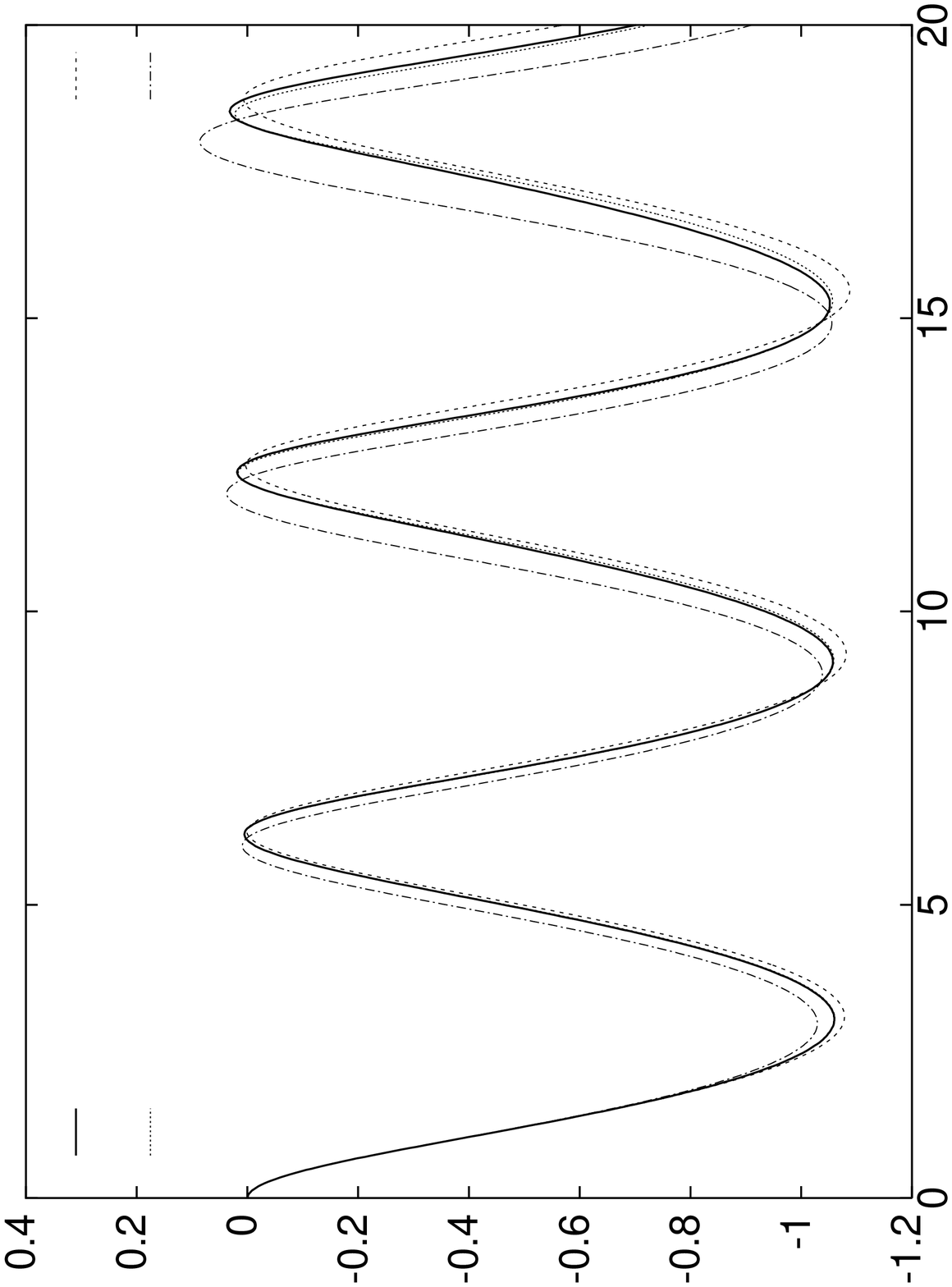}

   \put(79,-4){\normalsize  \boldmath$\theta$}
   \put(-8,55){\normalsize  \boldmath$\log(\gamma \xi)$}
   
   \put(31,97){\normalsize Exact}
   \put(31,89){\normalsize \boldmath$\log(\gamma \xi_E)$}

   \put(110,97){\normalsize  \boldmath$\log(\gamma \xi_M)$ }
   \put(110,89){\normalsize  \boldmath$\log(\gamma \xi_{MF})$ }

\end{picture}
\vspace{2mm}
\figcap{  Comparison of $\log( \gamma \xi )$ as a function of $\theta$ for various 
          analytical expressions of the correlation as derived in the text. The 
          parameters for all curves are $a=0.25$, $|\Delta|=0.5$, $n_b=0.15$ and $N=100$.
          The solid curve is the exact correlation length. In the large $N$
          limit all curves overlap completely. 
\label{corr_a025_Delta05_N100} }
\end{figure}

\vspace{0.5cm}   
\bc{
\subsection{The Mean Field Approximation}
}\ec

   We can also use a mean field approximation in order to get an 
   approximative and analytical expression 
   for the correlation length $\xi$. When the atoms have Poisson 
   distributed arrival times, the continuous master equation gives the
   following exact equation for the average photon occupation
   number \cite{ElmforsLS95}:

\be \label{dl_forv_x}
  \frac{1}{\gamma} \frac{d\langle x \rangle}{dt} = 
  -\left[ \langle x \rangle - \frac{n_b}{N} \right] + a \langle q_{n+1} \rangle - 
  b \langle q_n \rangle ~~.
\ee

   \noi
   By making use of the mean field approximation 
   $\langle q_n \rangle \approx q_{\langle n \rangle}= q(\langle x \rangle)$,
   the above equation simplifies considerably.
   The stationary solution, $\langle x \rangle = x_0$, of 
   \eq{dl_forv_x} is determined by a transcendent equation which, generally, can be 
   solved numerically only. However, with the approximation

\be \label{approx_q_MF}
  q(\langle x \rangle ) \approx  q(\langle x \rangle + 1/N) \approx q(\langle x \rangle + f/N)~~,  
\ee

   \noi
   the stationary solution $x_0$ has a simple parametric representation.
   To get a consistent expression for the correlation with \eq{corr_chi} 
   the weight factor $f$ has to be

\be
   f=\frac{a}{2a-1} ~~.
\ee

   \noi
   With the approximation \eq{approx_q_MF}, which clearly is very good for large $N$,
   \eq{dl_forv_x} reduces to

\be \label{dl_forv_x_approx}
  \frac{1}{\gamma} \frac{d\langle x \rangle}{dt} = 
  -\left[ \langle x \rangle - \frac{n_b}{N} \right] + (2a-1) \, 
  q(\langle x \rangle + f/N) ~~. 
\ee

   \noi
   The stationary solution $x_0$ of \eq{dl_forv_x_approx} then satisfies 
   the transcendental mean field equation

\be \label{x_0_lign}
  x_0 = \frac{n_b}{N}  + (2a-1) \, q(x_0+f/N) ~~. 
\ee

   \noi
   In the large $N$ limit,
   \eq{x_0_lign} reduces to  \eq{trans} after a multiplication by $x_0$.
   In contrast to the saddle-point equation \eq{trans}, 
   \eq{x_0_lign} therefore  
   always has the trivial $x_0=0$ solution in the large $N$ limit. 
   The general solution of \eq{x_0_lign} can be written in the 
   following parametric form:

\be \label{x_0_lign_param}
  x_0(\phi) = \frac{1}{2} \left( - h(\phi) + \sqrt{ h^2(\phi) + 4 \, g(\phi) } 
              \right)~~, 
\ee

   \noi where

\be \label{f}
  h(\phi) \equiv  \frac{f-n_b}{N} + \Delta^2 - (2a-1) \sin^2\phi~~,
\ee

\be \label{g}
  g(\phi) \equiv \frac{n_b \, f}{N^2} + \Delta^2 \frac{n_b}{N} + \frac{f}{N} \, 
          (2a-1) \sin^2\phi ~~,
\ee

  \noi and

\be \label{theta}
  \theta(\phi) =  \frac{\phi}{\sqrt{x_0(\phi) + f/N + \Delta^2} }~~.
\ee

   \noi
   For small perturbations around $x_0$, i.e. $\langle x \rangle = x_0 + \delta x$, we find
   the equation of motion

\be \label{delta_x_approx}
  \frac{1}{\gamma} \frac{d \, \delta x}{dt} = 
  -\left[ \, 1 - (2a-1) \, q^{ \, \prime}(x_0+f/N)  \, \right] \, \delta x~~, 
\ee

    \noi where

\bea \label{q_prime_x}
  q^{ \, \prime}(x) \equiv \frac{d \, q(x)}{dx}  
                  &=&  
                  \frac{\Delta^2}{(x+\Delta^2)^2} \, \sin^2( \theta\sqrt{x+\Delta^2})
                  \\ && \nonumber
                  + \, \, \,
                  \frac{x}{(x+\Delta^2)^{3/2}} \, \theta \,  
                  \sin( \theta\sqrt{x+\Delta^2}) \, 
                  \cos( \theta\sqrt{x+\Delta^2})~~.
\eea

   \noi
   In terms of $\phi$, \eq{q_prime_x} at $x=x_0(\phi)$ reads

\be
   q^{\,\prime}(\phi) = \frac{1}{(x_0(\phi)+f/N+\Delta^2)^2} 
                      \left[ \, \Delta^2 \sin^2\phi + x_0(\phi) \, \phi 
                      \sin \phi \cos \phi \, \right]~~.
\ee

   \noi  
   From \eq{delta_x_approx} we immediately see that the correlation 
   is given by

\be \label{xi_phi}
  \gamma \xi \simeq \gamma \xi_{MF}(\phi)  =  \frac{1}{1 - (2a-1) \, q^{ \, \prime}(\phi)}~~.
\ee

   \noi
   Notice that $\gamma \xi_{MF}$ is here a function of the parameter $\phi$,
   in contrast to \eq{corr_lambda} and \eq{corr_approx_lambda}.

   When the micromaser is not detuned, \eq{xi_phi} 
   reduces to

\be \label{corr_Delta0}
   \gamma \xi_{MF}(\phi) =   \frac{1}{1  - \displaystyle{
                       (2a-1)\frac{\phi \sin \phi \cos \phi}
                       {(2a-1)\sin^2\phi \, + \, \frac{n_b+f}{N}} } }~~.
\ee

   \noi
   The peak at $\theta=\theta_{0}^*$ corresponds to

\be \label{phi_0_ja}
  \phi = \phi_0^* \approx \left( \frac{3\, (n_b+f)}{N \, (2a-1)} 
                              \right)^{1/4}~~.
\ee

   \noi
   Substituting \eq{phi_0_ja} into equation \eq{corr_Delta0}
   the correlation is

\be
  \gamma \xi_{MF}(\phi_{0}^*) = \frac{2a-1}{2} \sqrt{ \frac{3 \, N}{a + n_b(2a-1)}   }~~,
\ee

   \noi
   for some $a$ the interval $1/2 \leq a \leq 1$.
   This correlation is in agreement with \eq{corr_chi}.

   In the thermal phase, where $x_0=0$ in the large $N$ limit,
   \eq{xi_phi} reduces to

\be \label{xi_phi_x0}
  \gamma \xi_{MF}(\theta) =  \frac{1}{1 - (2a-1) \, \theta_{eff}^2}~~,
\ee

   \noi
   in agreement with \eq{corr_termisk_master} and \eq{corr_lambda_termisk}.

\vspace{0.5cm}   
\bc{
\subsection{Statistical Barrier Penetration}
\label{tunnel_KAP}
}\ec

   As we have seen above, the correlation length
   exhibits large peaks for the pump parameters $\theta_0^*$,
   $\theta^*_{k k+1}$ and/or 
   $\theta^*_{tk}$. The correlation grows as $\sqrt{N}$ at $\theta = \theta^{*}_{0}$.
   At  $\theta^*_{k k+1}$ and $\theta^*_{tk}$ the large $N$ dependence is, however, different.
   At these values of the pump parameter 
   there is instead a competition between two neighbouring minima 
   of the effective potential $V_0(x)$ corresponding to, say, 
   $x=x_0$ and $x=x_2$. A barrier, corresponding to
   a local maximum of $V_0(x)$ at, say,  $x=x_1$, then separates these two 
   minima. Using the technique of \cite{ElmforsLS95} it can be
   shown that the peak in the correlation length close to $\theta = \theta_{k k+1}^*$ is 
   described by the expression

\be \label{xi_exp_N_max}
   \gamma \xi \simeq \frac{2\pi}
                { [ \, x_1(1+n_b) + b \, q(x_1) \, ] \,\sqrt{-V_0^{\prime \prime}(x_1)} \, \, f(x_0,x_1,x_2)} ~~,
\ee

  \noi where we have defined

\be
 f(x_0,x_1,x_2) = \sqrt{V_0^{\prime \prime}(x_0)} \, 
                        e^{-N[ V_0(x_1) - V_0(x_0)]} + 
                        \sqrt{V_0^{\prime \prime}(x_2) } \, 
                        e^{-N[ V_0(x_1) - V_0(x_2)]} ~~.
\ee

   \noi
   We therefore conclude that

\be 
  \gamma \xi \simeq e^{ N \Delta V_0} ~~,
\ee

    \noi 
    where $\Delta V_0$ is the smallest potential barrier between the two competing minima 
    of the effective potential $V_0(x)$. This equation shows explicitly that the correlation 
    length grows exponentially with $N$. 
    Here we observe that $V_0(\theta)$, and hence also $\Delta V_0$, 
    is proportional to the combination $a+n_b(2a-1)$ (see \eq{V_min}).
    Since the $\gamma \xi$-peaks grow exponentially 
    with $N$ they cannot be described by \eq{DL} with a finite power expansion of $q(x)$. 
        The first-order transitions at pump parameters corresponding to $\theta = \theta_{tk}^*$
    can be treated in a similar fashion.

\vspace{0.5cm}   
\bc{
\subsection{Discussion}
}\ec

   In Figs. \ref{corr_a1_Delta05_N100}, \ref{corr_a075_Delta05_N100}
   and \ref{corr_a025_Delta05_N100} we show
   the different analytical expressions for $\gamma \xi$ as given by Eqs. 
   (\ref{corr_lambda}), (\ref{corr_approx_lambda}) and (\ref{xi_phi}). 
   We have also plotted the exact correlation $\gamma \xi$.
   For sufficiently large $N$, all the three expressions are good approximations
   to the correlation length around the first peak (see Figs. \ref{corr_a1_Delta05_N100} 
   and \ref{corr_a075_Delta05_N100}). We also observe that the mean field
   approximation is a better approximation than \eq{corr_approx_lambda}.
   Moreover, the mean field approximation as well as the the assumption 
   $\theta \sqrt{x(\theta) + \Delta^2}~\ll~1$ 
   break down for pump parameters beyond the first peak of the correlation length. Hence, neither 
   \eq{corr_approx_lambda} nor \eq{xi_phi} is a good approximation for $\theta$'s above the
   first peak (see e.g. \fig{corr_a075_Delta05_N100}).
   \eq{xi_phi}, however, describes some qualitative features of the correlation length
   also for values of $\theta$ above the first peak, even though \eq{xi_phi} then gives
   numerical values which are quantitatively wrong.
	We finally notice that all curves in \fig{corr_a025_Delta05_N100} 
   overlap in the large $N$ limit.

\vspace{1cm}
%
\bc{
\section{TRAPPING STATES}
\label{trapping_Sec}
}\ec

   The stationary photon probability distribution  \eq{p_n_eksakt} has a special property when $n_b=0$ and 
   $q_m=0$. We then have that ${\bar p}_n=0$ for all $n \geq m$. The cavity then  
   cannot be pumped above $m$ by photon emission from the pump atoms. The micromaser 
   is then said to be in a trapping state \cite{Filipowicz86,Fili&Java&Meystre}-\cite{Slosser&Meystre_II}. Actually, such trapping states have recently been observed in the stationary state of the micromaser system \cite{W&V&H&W}.
For a given $\Delta$ and $k=1,2,3,...,$ we now define the function

\be \label{theta_trapp}
   \theta_k(x) = \frac{k \pi}{\sqrt{x + \Delta^2}} ~~,
\ee

   \noi
   or equivalently 

\be \label{theta_trapp_X}
   x_k(\theta) = \frac{(k \pi)^2}{\theta^2} - \Delta^2~~.
\ee

  \noi 
  For a given $N$, trapping states then occur at the pump parameters

\be
  \theta_{mk}^{tr} = \theta_k(x_{tr}) ~~,
\ee

  \noi
  where $x_{tr}=m/N$ and $m=0,1,2,3,...$~.

\begin{figure}[t]
\unitlength=1mm
\begin{picture}(100,80)(0,0)

\includegraphics{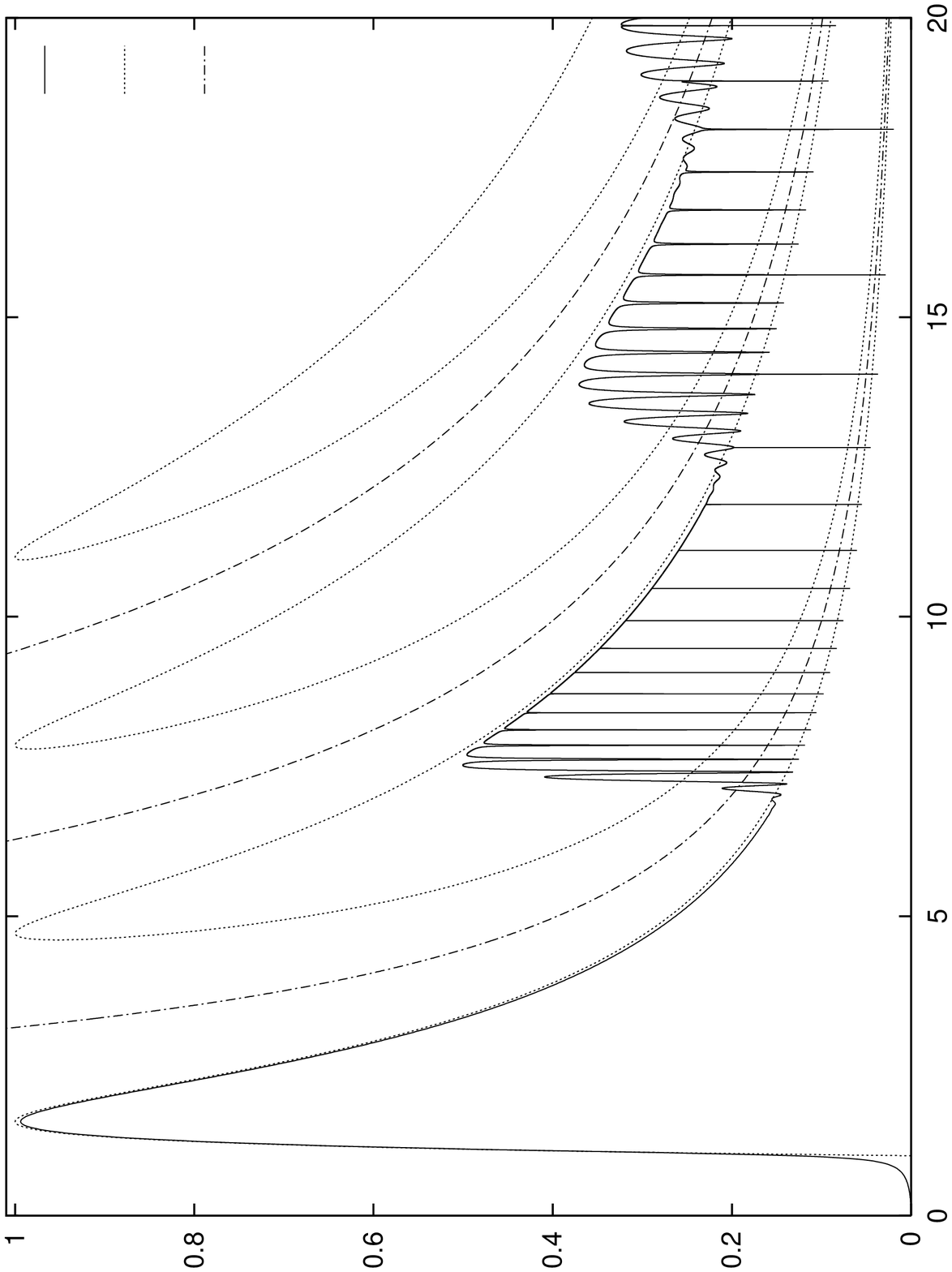}      

   \put(79,-4){\normalsize  \boldmath$\theta$}

   \put(122.4,98.5){\normalsize  \boldmath$\langle x \rangle$}
   \put(110,89.5){\normalsize  Mean field }
   \put(118,82){\normalsize \boldmath$x_k(\theta)$}

\end{picture}
\vspace{2mm}
\figcap{ The solid line shows the order parameter $\langle x \rangle = \langle n \rangle /N$ as a 
         function of $\theta$ when  $n_b=0,$ $a=1$, $\Delta=0$ and $N=100$. As a guide to the eye
         we have also plotted the mean field solution (see \eq{x_phi} and \eq{theta_phi}). 
         The dashed-dotted curve is $x_k(\theta)$ (see \eq{theta_trapp_X}).
         For a given $k\geq1$, this curve lies between the mean field solutions corresponding 
         to the branches $k$ and $k-1$.
\label{x_sfa_theta_trapp_fig} }
\end{figure}

The effect of trapping states on the order parameter $\langle x \rangle$ is illustrated in \fig{x_sfa_theta_trapp_fig} with an atomic flux parameter $N=100$. We then clearly 
   observe dips in the order parameter $\langle x \rangle$ due to the presence of trapping states. 
   The observed structure of the dips in $\langle x \rangle$ can be explained as follows. 
   For a given $k\geq1$, one can prove that the curve described by \eq{theta_trapp_X}
   lies between the mean field solution corresponding to branch $k$ 
   and $k-1$ (c.f.  \fig{x_sfa_theta_trapp_fig}). Because of this fact, and since $\langle n \rangle \leq m$, 
   the trapping-dip at $\theta^{tr}_{mk}$ reaches down to (in fact, just below) the mean field curve 
   corresponding to branch $k-1$ (c.f. \fig{x_sfa_theta_trapp_fig}).

In general, for a given $k$, trapping states  such that the
   value of $m/N$ is larger than the order parameter $\langle x \rangle$ have a minor effect on the order parameter.
When $\theta \lesssim \theta_{01}^{*}$ (or $\theta_{t1}^*$) the presence of trapping states therefore have little effect on the 
   order parameter (see e.g. \fig{x_sfa_theta_trapp_fig})
  since the curve described by \eq{theta_trapp_X}
   lies above the order parameter $\langle x \rangle$.

   It is easy to realize that trapping effects become less significant when the system 
   is detuned. In this case we see from \eq{x_phi} and \eq{theta_trapp_X} that $x_k(\theta)$ and 
   the mean field solution is just reduced by the same amount $\Delta^2$.   
   Numerical studies also show that
  the micromaser phase transitions occur essentially at values of the pump parameter $\theta$ almost 
   independent of the value of the detuning $\Delta^2 \leq 1$.
   Due to 
   these facts the order parameter $\langle x \rangle$, $x_k(\theta)$ in \eq{theta_trapp_X} as well as the 
   mean field solution will intersect with a shifted horizontal $\theta$-axe and hence  trapping 
   effects become less important.   In passing
 we notice that, for a given value of the detuning $\Delta$, trapping effects cannot
occur for 
$\theta \leq \theta^{tr}_{min} \equiv \pi/\sqrt{1+\Delta^2}$.
 This is due to the fact that the order parameter corresponding to the mean field solution (see \eq{x_phi} and \eq{theta_phi}) can never exceed unity.

   As the atomic flux $N$ increases, the dips in $\langle x \rangle$ due to trapping states become denser as a function of the pump parameter 
$\theta$ \cite{Meystre&Rempe&Walther}. 
In the large $N$ limit and for $n_b = 0$, $\langle x \rangle$ therefore ceases to be
an appropriate order parameter and the system appears to be frustrated.
Since the observable $\langle x \rangle$ then varies rapidly it is natural to ask
   whether it is well defined at all. When the system is in a maser phase 
   and when $N$ is sufficiently large, 
   numerical studies of the standard deviation 
   $\Delta x = \sqrt{ \langle x^2 \rangle - \langle x\rangle^2}$ as a function of $\theta$
   show that $\Delta x$ is much smaller than the value of 
   $\langle x \rangle$ itself for all possible
   values of $\theta$, except for values of
   $\theta$ near the phase transitions $\theta = \theta^*_{kk+1}$ and/or $\theta=\theta^*_{tk}$. 
   Hence, the observable $\langle x \rangle$  is actually well defined except near these particular pump parameters.

\begin{figure}[t]
\unitlength=1mm
\begin{picture}(100,80)(0,0)

\includegraphics{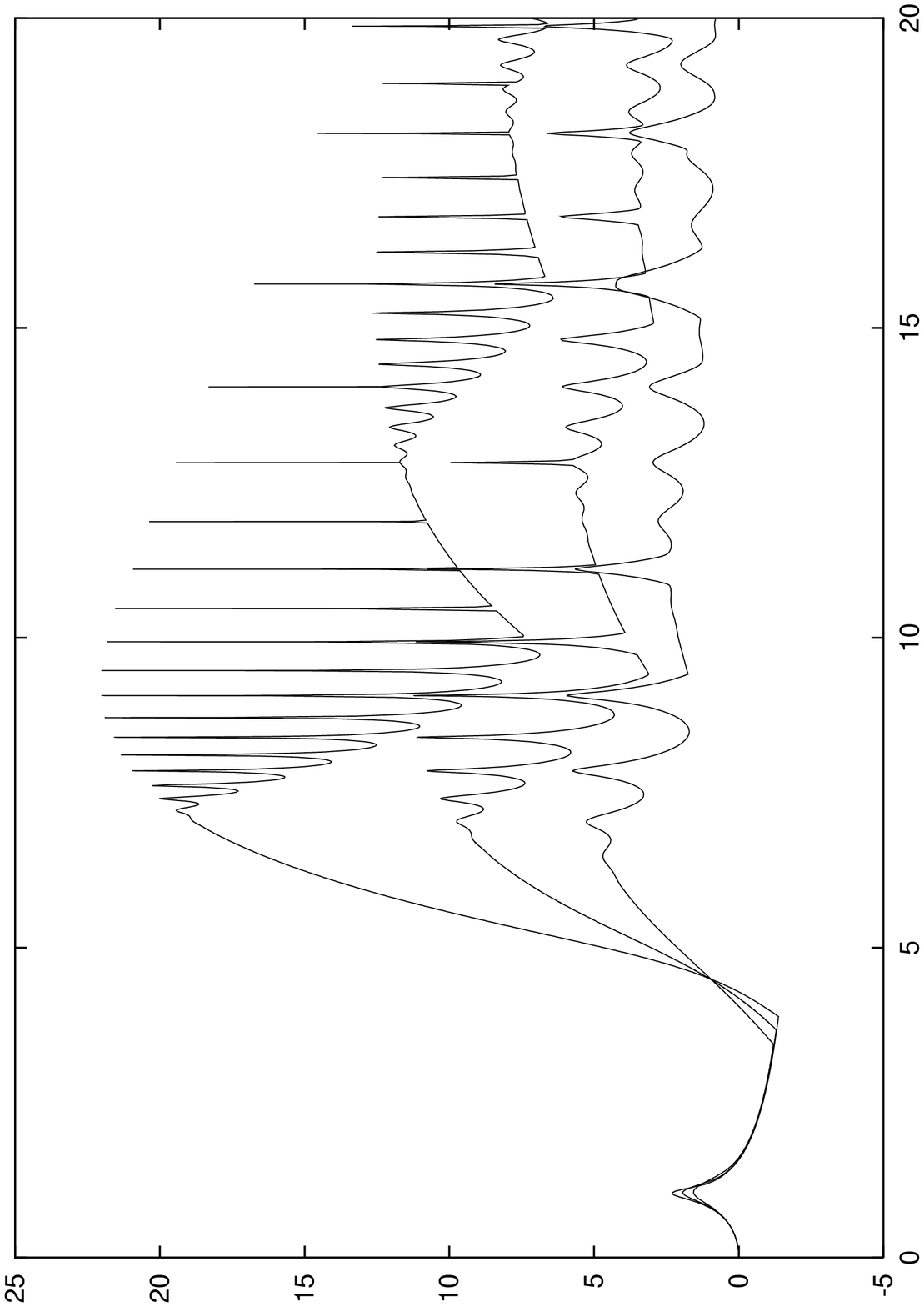}

   \put(-8,55){\normalsize  \boldmath$\log( \gamma \xi )$}
   \put(77,-4){\normalsize  \boldmath$\theta$}

   \put(46,64){\vector(1,0){6}}
   \put(26,63){\normalsize  \boldmath$N=100$}

   \put(44,49){\vector(1,0){10}}
   \put(25,48){\normalsize  \boldmath$N=50$}

   \put(52,27){\vector(0,1){9}}
   \put(44.5,22){\normalsize  \boldmath$N=25$}

\end{picture}
\vspace{2mm}
\figcap{ The logarithm of $\gamma \xi$ as a function of $\theta$ when $N=25$, $N=50$ 
         and $N=100$ and where $n_b =0$, $a=1$ and $\Delta = 0$.
\label{xi_N_fig} }
\end{figure}

   In \fig{xi_N_fig} we have studied the correlation length for different values of $N$. 
   The effect of trapping states becomes more pronounced when $N$ increases.
   We observe that the peaks in the correlation length corresponding to trapping states get denser for large values of $N$.
   Because of the same reasons as discussed earlier in this section,
   the effect of trapping states on the correlation length become less
   significant when the system is detuned.

\begin{figure}[t]
\unitlength=1mm
\begin{picture}(100,80)(0,0)

\includegraphics{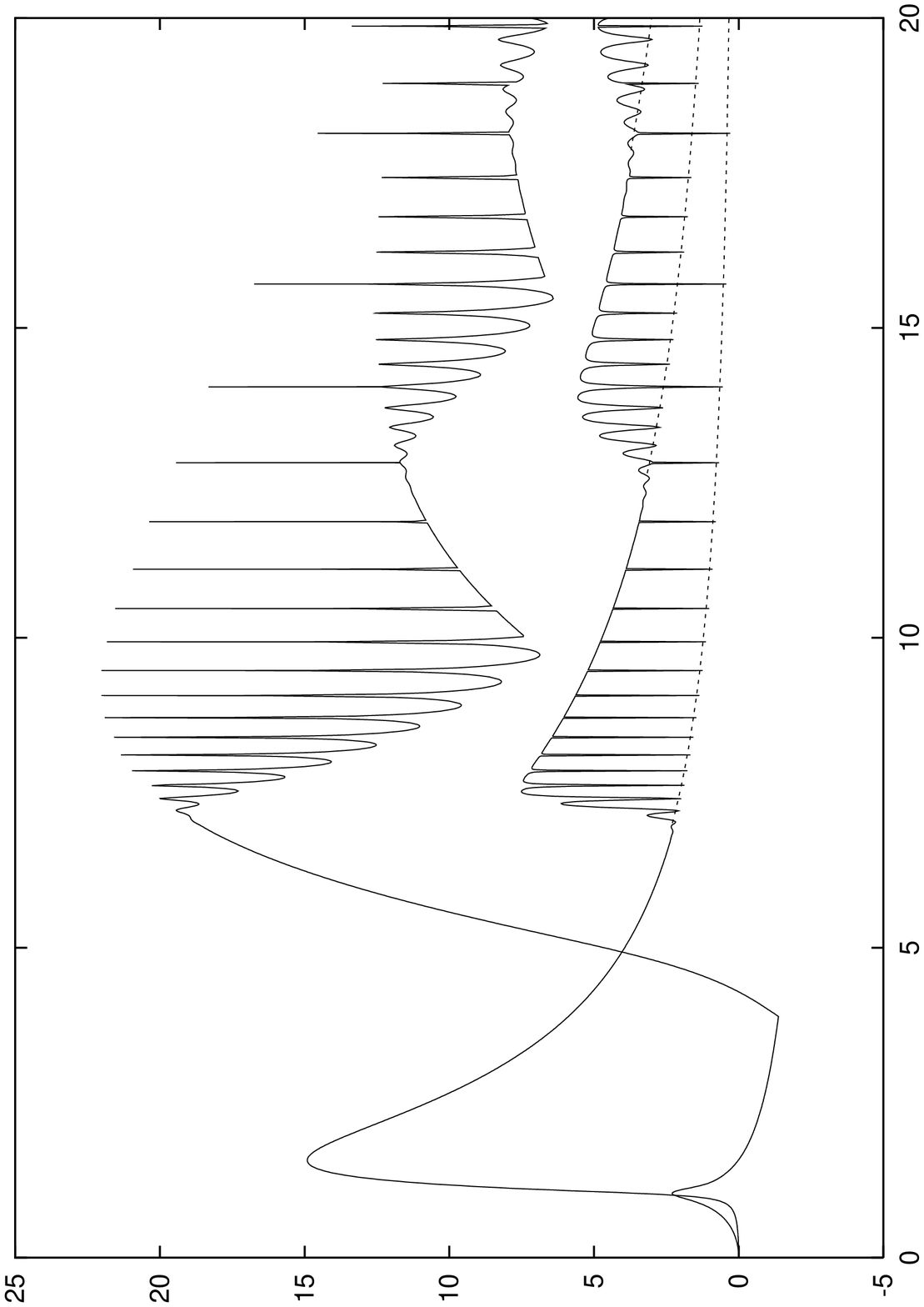}      

   \put(77,-4){\normalsize  \boldmath$\theta$}

   \put(24,75.3){\vector(0,-1){5}}
   \put(26.5,75.4){\oval(5,5)[tl]}
   \put(28.8,77.5){\normalsize  \boldmath$15 \langle x \rangle$}

   \put(43,22){\vector(0,1){5}}
   \put(45.5,22.5){\oval(5,5)[bl]}
   \put(47.5,19.3){\normalsize  \boldmath$\log( \gamma \xi )$}

\end{picture}
\vspace{2mm}
\figcap{ The logarithm of $\gamma \xi$  and $15 \langle x \rangle$
         as a function of $\theta$ when the parameters are as in  \fig{x_sfa_theta_trapp_fig}. 
         There is no visible difference between the $\theta$-positions
         of the correlation trapping-peaks and the $\theta$-positions of the order parameter 
         trapping-dips.
         The dotted lines are the mean field solution \eq{x_phi} and \eq{theta_phi}
         for branch $k=0$, $k=1$ and $k=2$ for some limited interval of $\phi$.
\label{xi_og_x_fig} }
\end{figure}

    As mentioned in Section~\ref{corr_KAP}, the correlation length 
    reaches a  maximal value when the potential 
    barrier between the two competing minima of the effective potential $V_0(x)$ is at its lowest 
    value. 
    If the system is not in a trapping state,
    the probability distribution ${\bar p}_n$ then 
    essentially consists of two Gaussian peaks with heights which
    are of the same order of magnitude. 
    Furthermore, due to the pre-factor $w(x)$ in ${\bar p}(x)$ (c.f. \eq{p_x}), 
    a micromaser system which is close
    to a trapping state will also have a probability distribution which 
    essentially consists of two Gaussian peaks of the same order of magnitude, 
    even though the potential barrier between the two competing 
    minima of the effective potential $V_0(x)$ will then be smaller.
    Hence, the correlation length again reaches a large value. Exactly at a trapping state, however,
    the probability distribution consists essentially of  one 
    dominating Gaussian peak in the large $N$ limit. This means that we 
    should expect the correlation length to become large close
    to a trapping state however not exactly
    at the occurence of such a state. Numerical studies are in accordance with these observations. As illustrated in   
    \fig{xi_og_x_fig}, there is, furthermore, no visible difference between the $\theta$-positions
    of the peaks of $\gamma \xi$ and the $\theta$-positions of the dips of the $\langle x \rangle$.
    This is not an  obvious fact since $\xi$ measures long-time features of the micromaser system but $\langle x \rangle$ is 
    an instantaneous observable.
   As the number of branches increases, that is, the number of local maxima in ${\bar p}(x)$
   increases, the connection between the heights of the peaks of $\xi$ and the depths of the
   dips of $\langle x \rangle$ becomes increasingly more complex.

    As in Section \ref{corr_KAP}, let $\lambda_n$ denote the eigenvalues of the matrix $L$. 
    Introducing the cumulative probability

\be
  P_n = \sum_{m=0}^{n-1} {\bar p}_n ~~,
\ee

  \noi
  it was then shown in  Ref.\cite{ElmforsLS95} that

\be \label{corr_sum_KONV}
  \sum_{n=1}^\infty \left ( \frac{1}{\lambda_n} - \frac{1}{\lambda_n^0} \right ) 
   = \sum_{n=1}^\infty \left (  \frac{ P_n( 1 - P_n) }{ [(1+n_b)n +N b q_n] \, {\bar p}_n }
                          -  \frac{1 - [n_b/(1+n_b)]^n}{n} \right ) ~~,
\ee

  \noi
  where $\lambda_n^0 \simeq n$ is an eigenvalue  corresponding to the thermal
 equilibrium distribution. 
  In the large $N$ limit the left-hand side can be approximated by $\gamma \xi -1$.
  Hence, \eq{corr_sum_KONV} reads

\be \label{corr_sum_APPROX}
  \gamma \xi \approx  1 +
  \sum_{n=1}^\infty \left (  \frac{ P_n( 1 - P_n) }{ [(1+n_b)n +N b q_n] \, {\bar p}_n }
                          -  \frac{1 - [n_b/(1+n_b)]^n}{n} \right ) ~~.
\ee

  \noi
   When the parameters are as in \fig{x_sfa_theta_trapp_fig} we find that, except for a small 
  interval $2\lesssim \theta \lesssim 4$, the sum-rule prediction  
  \eq{corr_sum_APPROX} is in surprisingly good agreement with the exact correlation length,
  even at trapping states. We have no deep explanation of this numerical observation.

\vspace{1cm}
\bc{
\section{FINAL REMARKS}
\label{final_KAP} 
}\ec
\vspace{0.5cm}
We have studied the micromaser phase transitions at non-zero detuning and for
pump atoms prepared in a diagonal statistical mixture. New novel features
of the micromaser system then emerge as plateaux in the order parameter 
   $\langle x \rangle$ as a function of the atomic density matrix as well as
a twinkling behaviour at non-zero detuning. 
By introducing fluctuations in the pump-parameter $\theta$ one can decrease the
signal-to-noise ratio in observables like ${\cal P}(+)$ and ${\cal P}(+,+,t)$ \cite{Filipowicz86,ElmforsLS95,seoul&99}. 
Elsewhere we will investigate this feature of
noise synchronization in more detail, 
also when the detuning $\Delta$ parameter is a random variable. It turns out that
fluctuations in $\Delta$ also lead to features similar to fluctuations in $\theta$,
as was already anticipated in Ref.\cite{Filipowicz86}. 
Fluctuations in the atomic density matrix parameter $a$ lead, however, not to an
output signal of the micromaser with less fluctuations.

\vspace{3mm}
\begin{center}
{\bf ACKNOWLEDGEMENT}
\end{center}
%

   The authors wish to thank A. De R\'{u}jula and the members of the TH-division 
   at CERN for the warm hospitality while 
   the present work was completed.
   The research has been supported in part  by the Research Council of
   Norway under contract no. 118948/410. 
   One of the authors (B.-S.S.) wishes to thank H. Walther for discussions
	and useful correspondence.

\vspace{3mm}

%
 
%
\end{document}